\documentclass[showpacs,amsmath,amssymb,pra,twocolumn,superscriptaddress,notitlepage,longbibliography]{revtex4-1}

\usepackage[dvips]{graphicx} 
\usepackage{amsmath,amssymb,amsthm,mathrsfs,amsfonts,dsfont}
\usepackage{enumerate}
\usepackage{epsfig}
\usepackage{subcaption}
\usepackage{xcolor}
\usepackage{amsthm}
\usepackage{braket}
\usepackage{epstopdf}
\usepackage{multirow}
\usepackage{array}     
\usepackage{comment}
\usepackage[normalem]{ulem}
\usepackage{enumitem} 
\usepackage[colorlinks = true]{hyperref}
\usepackage{float}
\newcommand{\tr}{\mathrm{Tr}}

\newtheorem{theorem}{Theorem}
\newtheorem{lemma}{Lemma}

\newtheorem{observation}{Observation}

\graphicspath{{./figure/}}

\begin{document}

\title{Reference-frame-independent design of phase-matching quantum key distribution}

\author{Anran Jin}
\affiliation{Electrical Engineering Division, Department of Engineering, University of Cambridge, CAPE Building 9 JJ Thomson Avenue, CB3 0FA Cambridge, UK}

\author{Pei Zeng}
\affiliation{Center for Quantum Information, Institute for Interdisciplinary Information Sciences, Tsinghua University, Beijing 100084, China}

\author{Richard V. Penty}
\affiliation{Electrical Engineering Division, Department of Engineering, University of Cambridge, CAPE Building 9 JJ Thomson Avenue, CB3 0FA Cambridge, UK}

\author{Xiongfeng Ma}
\email{xma@tsinghua.edu.cn}
\affiliation{Center for Quantum Information, Institute for Interdisciplinary Information Sciences, Tsinghua University, Beijing 100084, China}

\begin{abstract}
The recently proposed phase-matching quantum key distribution offers means to overcome the linear key rate-transmittance bound. Since the key information is encoded onto the phases of coherent states, the misalignment between the two remote reference frames would yield errors and significantly degrade the key generation rate from the ideal case. In this work, we propose a reference-frame-independent design of phase-matching quantum key distribution by introducing high-dimensional key encoding space. With encoded phases spanning the unit circle, the error statistics at arbitrary fixed phase reference difference can be recovered and treated separately, from which the misalignment angle can be identified. By naturally extending the binary encoding symmetry and complementarity to high dimensions, we present a security proof of this high-dimensional phase-matching quantum key distribution and demonstrate with simulation that a 17-dimensional protocol is completely immune to any degree of fixed misalignment and robust to slow phase fluctuations. We expect the high-dimensional protocol to be a practical reference-frame-independent design for general phase-encoding schemes where high-dimensional encoding is relatively easy to implement. 
\end{abstract}

\maketitle
\section{INTRODUCTION}
Quantum key distribution (QKD) resorts to quantum systems to distribute private and random keys between two distant parties. Moreover, the privacy does not rely on any computational assumption as in the classical key distributions, nor the randomness derived from any pseudo random number generations. In fact, the perfect privacy and randomness are intrinsic in quantum systems, as the security of QKD can be proved by reducing it to the distillation of entangled quantum states \cite{lo1999Unconditional,shor2000Simple,koashi2009simple}. 

The traditional QKD protocols are essentially point to point, where one party transmits quantum states according to the classical keys and the other party receives and measures the quantum states to distinguish the corresponding classical keys \cite{bennett1984quantum, bennett1992quantum}. Under this formalism, a linear bound can be placed on the relation between channel transmittance and key generation rate \cite{takeoka2014fundamental,pirandola2017fundamental}. This is reasonable, since only when the encoded states are transmitted through the channel can they be detected and used to generate raw keys. In 2012, the measurement-device-independent quantum key distribution (MDI QKD) was presented \cite{lo2012measurement}, originally intended to remove all of the detection loopholes. Yet moreover, MDI QKD employs a setup that differs from the traditional point-to-point scheme, in the sense that an untrusted third party is in between the two communicating parties. Through the claimed Bell-state measurements of the third party, the communicating parties can entangle their qubits through entanglement distillation.

The original MDI QKD, although having an untrusted relay in between, still cannot break the linear key rate-transmittance bound. This is essentially because MDI QKD encodes entanglements in the degree of freedom of photons, for instance polarization. In this way, two photons need to be detected for one successful detection, which yields one bit raw key. Since the third party who makes detection is in the middle of the two communicating parties, the detection rate of each photon is the square root of the total channel transmittance, and hence the total detection rate, which requires two photons, still scales linearly with the channel transmittance. 

The breakthrough was made in the recent work of twin-field quantum key distribution (TF QKD) \cite{lucamarini2018overcoming}, which opens the possibility of phase-encoding MDI QKD protocols. In this MDI scheme with single-photon interference \cite{Pfleegor1967Interfer}, a successful detection requires in total one photon from the two sides, saving the detection compared to the original MDI QKD schemes. As a result, TF QKD improves the key rate-transmittance bound from linear to square root. Afterwards, variants of TF QKD and their rigorous security proofs have been presented~\cite{Ma2018phase, lin2018simple, curty2019simple}. Among these works, a scheme named phase-matching quantum key distribution (PM QKD) \cite{Ma2018phase} encodes the key bits in the phase of coherent states and removes the need of basis switching, which has been demonstrated in a 502-km fiber-based experiment~\cite{Fang2020implementation}.

Despite their ability to enhance the key-rate performance in theory, these phase-encoding MDI QKD protocols are more challenging when it comes to the experimental implementation, due to the optical-mode quadrature reference mismatch between the two parties from the laser sources and optical channels. The reference mismatch is in fact a problem for general QKD systems, as in the polarization-based BB84 protocol mismatch between polarization axes gives a maximally tolerable misalignment error rate of 11\% \cite{shor2000Simple}. In Ref.~\cite{Zeng2019Symmetryprotected}, the feasibility of PM QKD under 13\% misalignment error rate was demonstrated, yet with severely discounted achievable secure key rate. 
In practice, phase-locking techniques can be employed to fix the phase reference \cite{Santarelli1994Het}, but the experimental challenges and the cost are considerable. The difficulty of phase locking is also reflected in the recent experimental demonstrations, which are either simple demonstration with local settings~\cite{minder2019experimental,zhong2020proofofprinciple} or highly demanding experiments with advanced technologies such as lasers with narrow linewidth of 1Hz and active phase feedback controls~\cite{wang2019beating}, ultrastable cavity and time-frequency transfer locking~\cite{liu2019experimental}, and laser-injection techniques~\cite{Fang2020implementation}. We refer to Ref.~\cite{Mao2021recent} for a detailed review on different variants of PM QKD and TF QKD protocols and the advances on the experimental techniques. Phase postcompensation is another feasible approach \cite{Ma2012alternative,Ma2018phase,Zeng2019Symmetryprotected}, where extra phase randomization is introduced and the experimental data with aligned phase slices are postselected afterwards. If the phase mismatch is relatively fixed, the data with aligned phase slices will be suitable for key generation.
This approach is again experimentally complicated and requires great amount of data for phase estimations. 

We thus call for the reference-frame-independent design \cite{Laing2010RFI,Lee2020RFIMDI} of PM QKD to cope with fixed or slowly fluctuating phase misalignment, completely controlled by the adversary in the worst case, without overcomplicating the experimental setups.

If we look at the essence of the phase postcompensation, the discrete randomization in fact expands the key space from two dimensions to high dimension. After the detection stage, the key space is reduced back to two-dimensional through postselection of matching phases. The variation in the key-space dimension complicates the protocol. Naturally, we can remove the postselection stage by implementing high-dimensional key space from the beginning. The potential of high-dimensional protocols against channel errors is already demonstrated for prepare-and-measure protocols, where, in contrast with the conventional two-dimensional BB84 protocol which tolerates an error rate of 11\% , the four-dimensional BB84 protocol can tolerate up to 35.6\% \cite{Chau2005HDQKD}, and the 16-dimensional BB84 protocol can tolerate 45.4\% \cite{Chau2005HDQKD}. These results shine light on introducing high-dimensional PM QKD to combat errors introduced by misalignment.

In this work, we introduce the $d$-dimensional PM QKD protocol which encodes key information onto $d$ uniformly separated phase slices and matches phases via interference detection at an untrusted measurement site. By extending the encoding symmetry approach \cite{Zeng2019Symmetryprotected} to high dimensions, we present a security analysis of the high-dimensional PM QKD and demonstrate that it achieves reference-frame independence: it is completely immune to any degree of fixed-phase misalignment and robust to small phase fluctuation, where these disturbances are assumed to be controlled by the adversary. As the high-dimensional PM QKD employs the same setup as the two-dimensional PM QKD whilst removing the necessity of phase postcompensation, it is in fact a pragmatic approach to mitigate the effect of reference mismatch.

The rest of the paper is arranged as follows. In Section~\ref{2Protocol}, we present the high-dimensional PM QKD protocol and discuss its relations with the conventional two-dimensional PM QKD. In Section~\ref{3Proof}, we outline  the security proof of high-dimensional PM QKD. The proof is generalized from two-dimensional encoding symmetry \cite{Ma2018phase,Zeng2019Symmetryprotected} and phase-error correction \cite{lo1999Unconditional,shor2000Simple,koashi2009simple}, and we will justify the elements that cannot be generalized to high dimensions trivially. Finally, in Section~\ref{4Comparison}, we present the simulation performance of the 17-dimensional PM QKD and demonstrate its advantage over the two-dimensional PM QKD against both fixed-phase misalignment and small phase fluctuation. We justify the rationales behind this advancement.

\section{HIGH-DIMENSiONAL PM QKD PROTOCOLS}
\label{2Protocol}
\begin{figure}[htbp!] 
\centering    
\captionsetup{justification = raggedright}
\includegraphics[width=0.45\textwidth]{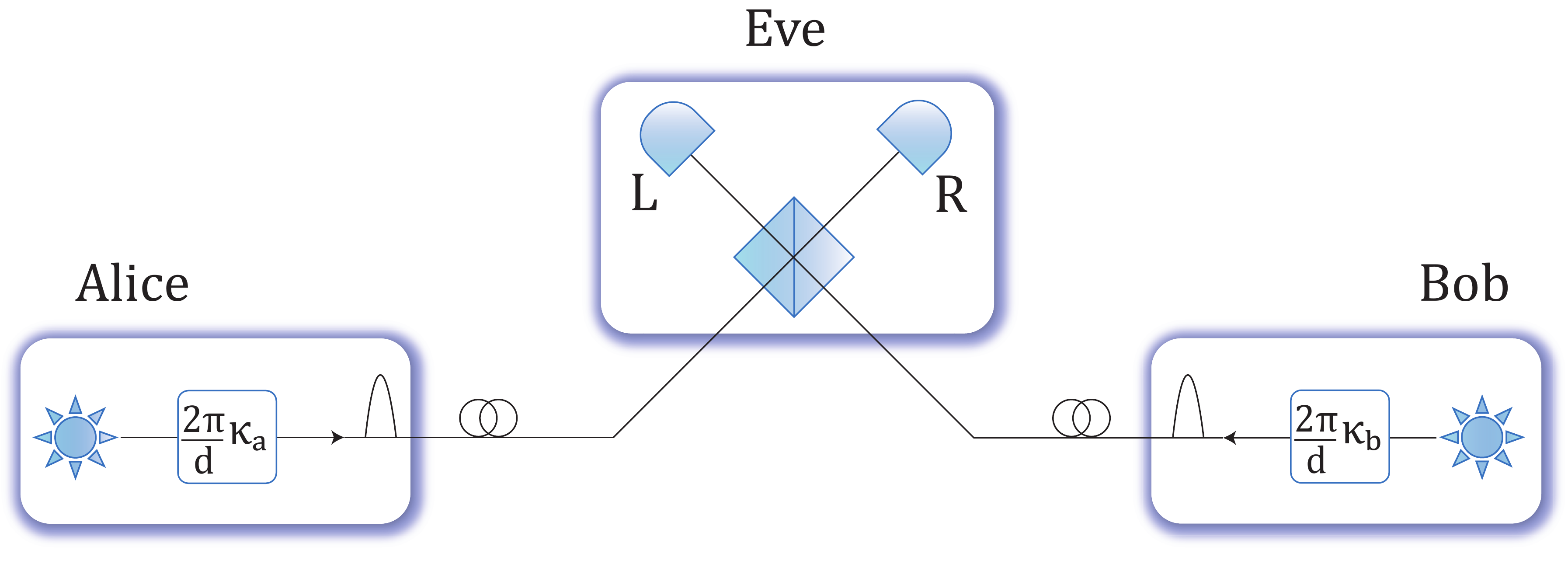}
\caption{Schematic diagram of the PM QKD protocol with $d$-dimensional encoding \cite{Ma2018phase}. Alice prepares the coherent state $\ket{\sqrt{\mu/2}~e^{i\frac{2\pi}{d} \kappa_a}}_A$, where $\kappa_a \in \{0,1,\cdots,d-1\}$. Similarly Bob prepares $\ket{\sqrt{\mu/2}~e^{i\frac{2\pi}{d} \kappa_b}}_B$. They send the two coherent states to interfere at an untrusted measurement site. Ideally, if the phase difference $\frac{2\pi}{d}|\kappa_a - \kappa_b| = 0$, the detector gives a single $L$ click. If $\frac{2\pi}{d}|\kappa_a - \kappa_b| = \pi$, the detector gives a single $R$ click.}
\label{PMQKD}
\end{figure}

We introduce the high-dimensional PM QKD protocol as the following, with Fig.~\ref{PMQKD} as a schematic diagram:

\textbf{\underline{High-dimensional PM QKD protocol}}
\begin{enumerate}
\item \textbf{Encoding}: Alice randomly generates a key ``dit" $\kappa_a$ from $\{0,1,\cdots,d-1\}$ and prepares the coherent state $\ket{\sqrt{\mu/2}~e^{i\frac{2\pi}{d} \kappa_a}}_A$. Similarly, Bob randomly picks $\kappa_b$ and prepares $\ket{\sqrt{\mu/2}~e^{i\frac{2\pi}{d} \kappa_b}}_B$.
\item \textbf{Measurement}: Alice and Bob send the two optical modes $AB$ to an untrusted party, Eve, who is supposed to perform interference measurement and announce the detection results: no click, double click, $L$ click or $R$ click. 
\item \textbf{Sifting}: After many rounds of quantum communications, Alice and Bob keep only the rounds with $L$ or $R$ click. They end up with two correlated $d$-dimensional strings.
\item \textbf{Parameter estimation}: From the raw data they retained, Alice and Bob estimate the security parameters and derive the secure key rate.
\item \textbf{Key generation}: Based on the parameter estimation results, Alice and Bob reconcile their raw strings by consuming certain secure keys. They then perform privacy amplification to extract the secure final keys from the reconciled keys.  
\end{enumerate}

This family of protocols is a direct generalization of the two-dimensional PM QKD \cite{Ma2018phase, Zeng2019Symmetryprotected} to $d$ dimension. The intuition of security is similar as the binary case: when Alice and Bob each send coherent states $\ket{\alpha e^{i\theta_a}}$ and $\ket{\alpha e^{i\theta_b}}$ to the interferometer, the device is highly likely to produce a single $L$ click only when $|\theta_a-\theta_b|\approx 0$, and a single $R$ click only when $|\theta_a-\theta_b|\approx \pi$. Hence, if they group the rounds with single $L$ clicks and $R$ clicks respectively, they would obtain a pair of correlated phase strings. They can then distill secure keys, respectively from the $L$-click group and the $R$-click group, and the total secure key length is the sum of that from the two groups \cite{GLLP2004,Ma2008PhD}. 
We note that similar protocols with discrete phase randomization are discussed in Refs.~\cite{CurrsLorenzo2021} and~\cite{Wang2020}. However, these protocols adopt binary encoding essentially; the discrete phase randomization is used for a tight parameter estimation. In contrast, the proposed high-dimensional PM QKD in this work utilizes the discrete random phases for a high-dimensional encoding.

\section{SECURITY OF HIGH-DIMENSIONAL PM QKD WITH ENCODING SYMMETRY}
\label{3Proof}
In this section, we sketch the security analysis of high-dimensional PM QKD protocols. A complete rigorous security proof is placed in the Appendices. Generally, the proof extends the binary symmetric encoding approach in Ref.~\cite{Zeng2019Symmetryprotected} to higher dimensions, which is discussed in Section~\ref{sec3A}, and concludes privacy through the phase-error correction approach in Ref.~\cite{koashi2009simple} in Section~\ref{sec3B}. In order to generalize the known results in two dimensions, we resort to the structure of finite field GF($d$) (see Appendix~\ref{AppendixA1}), which only exists when $d = p^r$ for some prime number $p$ and integer $r$. Hence, we confine the analysis to prime power dimensions. Due to a small caveat to be mentioned in Section~\ref{5Conclusion}, unless noticed (e.g., Section~\ref{sec3B}), we confine $d$ to prime numbers. We give the asymptotic key-rate formula for $d$-dimensional PM QKD with experimentally accessible parameters in Section~\ref{sec3C}.

\subsection{High-dimensional symmetric encoding protocol}
\label{sec3A}
We first consider the symmetric encoding property of the $d$-dimensional PM QKD \cite{Zeng2019Symmetryprotected}. In a $d$-dimensional symmetric encoding QKD, Alice and Bob start with a bipartite state $\rho_{AB}$. They independently generate a random key ``dit'' $\kappa_{a}$ and $\kappa_{b}$ from $\{0,1,\cdots,d-1\}$ and apply $U(\kappa_{a(b)}) := U^{\kappa_{a(b)}}$ to their subsystem $A$ and $B$ respectively, where $U^{d} = I$. Notice that in PM QKD, the encoding operator $U$ is the rotation operator
\begin{equation}
U = e^{i\frac{2\pi}{d} a^\dagger a},
\end{equation}
that rotates a coherent state by an angle of $2\pi/d$. The modulated state $\rho'_{AB}(\kappa_a,\kappa_b)$ can be written as
\begin{equation}
\rho'_{AB}(\kappa_a,\kappa_b) = [U_A(\kappa_a)\otimes U_B(\kappa_b)]\rho_{AB}[U_A(\kappa_a)\otimes U_B(\kappa_b)]^\dagger,
\end{equation}
which is then sent to the third party Eve who is supposed to make a joint measurement to determine $(\kappa_a - \kappa_b)\text{ mod } d$ and announce the result. Based on the announcements from Eve, Alice and Bob can modify their key dits to generate a pair of correlated key strings, with information reconciliation and privacy amplification to generate the final secure key. 

\begin{figure}[htbp!] 
\centering    
\captionsetup{justification = raggedright}
\includegraphics[width=0.45\textwidth]{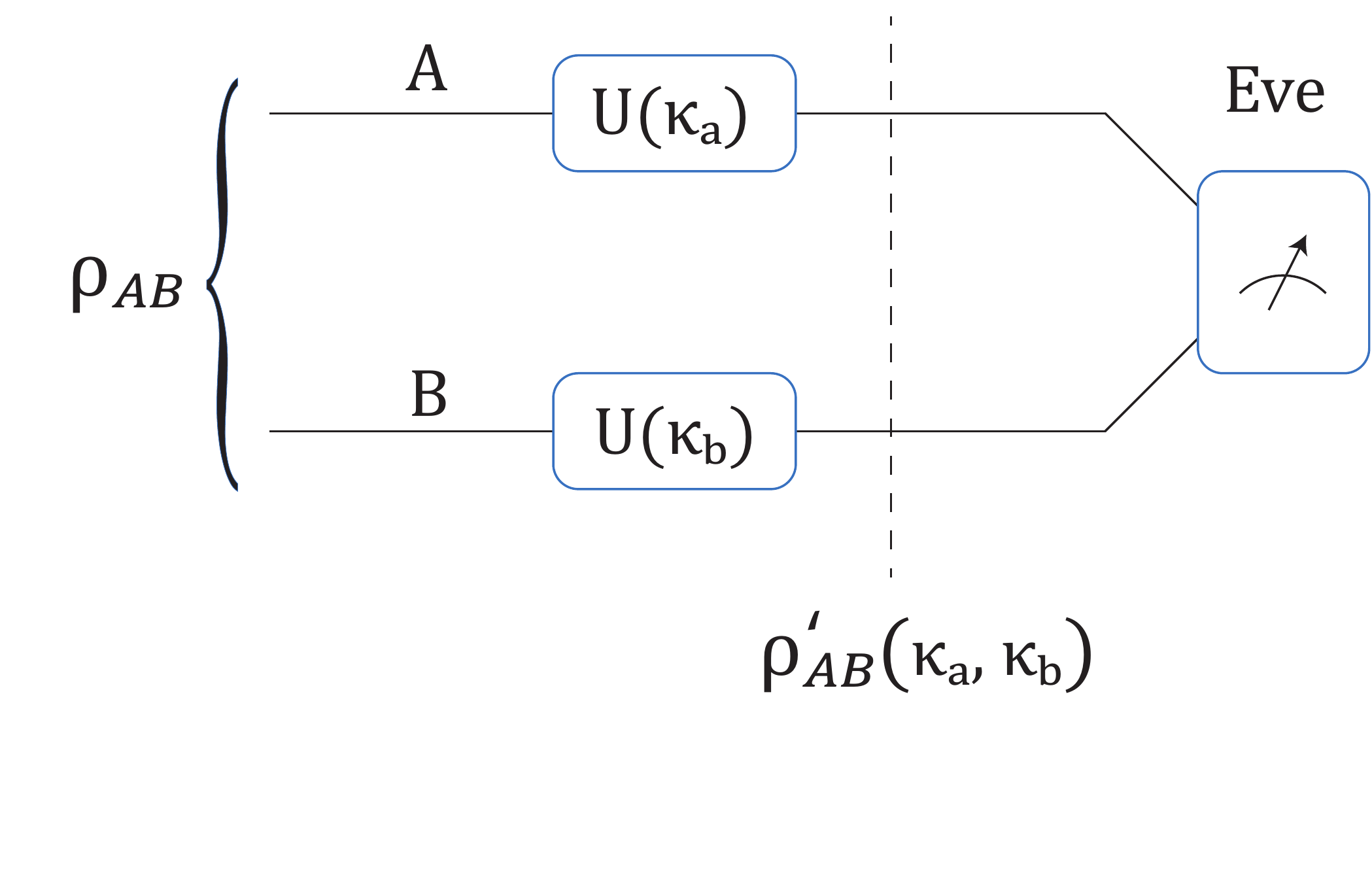}
\caption{ Schematic diagram of the $d$-dimensional symmetric encoding QKD, where $\kappa_a$, $\kappa_b$ take values in $\{0,1,\cdots,d-1\}$. Alice and Bob share the bipartite state $\rho_{AB}$ and each applies a unitary operation $U(\kappa) := U^\kappa$ according to their key values $\kappa_a$ and $\kappa_b$. The untrusted party Eve is supposed to announce the key difference $(\kappa_a - \kappa_b) \text{ mod } d$. The setup extends the binary symmetric encoding protocol in Ref. \cite{Zeng2019Symmetryprotected}.}
\end{figure}

A pure state $\ket{\psi}_{AB}$ on $AB$ is called an $l$-symmetric state, for $l$ in $\{0,1,\cdots,d-1\}$, if it is the $l$-th eigenstate of $U_A\otimes U_B$:
\begin{equation}
(U_A\otimes U_B)\ket{\psi}_{AB} = \gamma_d^l \ket{\psi}_{AB},
\end{equation}
where $\gamma_d = e^{i2\pi/d}$. For a mixture of $l$-symmetric states, $\rho_{AB} = \sum_j p_j \ket{\psi_l^{(j)}}\bra{\psi_l^{(j)}}$, we have
\begin{equation}
\rho'_{AB}(\kappa_a,\kappa_b) = [I_A \otimes U_B(\kappa_b - \kappa_a)] \rho_{AB}, 
\end{equation}
where the subtraction is under modulus $d$. Hence, the encoded mixture $l$-symmetric states are indistinguishable as long as the two key dits $\kappa_a$ and $\kappa_b$ differ by the same number. As a result, the raw key dit $\kappa_a$ is ``hidden'' in the encoded state $\rho'_{AB}(\kappa_a,\kappa_b)$ as long as the preshared state $\rho_{AB}$ is a mixture of pure parity states.      

To give a more rigorous argument, we resort to the entanglement-based symmetric encoding protocol, as shown in Fig.~\ref{EBSymmetric} below. In the entanglement-based protocol, Alice and Bob each hold an ancillary system $A'$ and $B'$ in the state $\ket{+}_d = \sum_{j=0}^{d-1}\ket{j}$. This serves as the control dit of the encoding operator $U$, i.e. we transfer the classical random encoding to a quantum control operation. Its equivalence with the prepare-and-measure symmetric encoding protocol follows if we move the final measurement prior to the control operation. For the entanglement-based protocol, when the input state $\rho_{AB}$ is an $l$-symmetric state, the following is shown in Appendix~\ref{Appendix3}:
\begin{observation}
\label{Theorem1}
In the entanglement-based symmetric encoding protocol, if the input state $\rho_{AB}$ is an $l$-symmetric state, then the $X$-basis measurement result on $A'$ can be determined by the $X$-basis measurement result on $B'$, as they sum to $l$ under modulus $d$ addition. In other words, Alice and Bob have no phase error between $A'$ and $B'$.
\end{observation}

Note that the $X$-basis complementary to the computational basis in a $d$-dimensional Hilbert space is defined as
\begin{equation}
\ket{\tilde{l}} := \frac{1}{\sqrt{d}} \sum_{j=0}^{d-1} \gamma_d^{-lj} \ket{j},
\end{equation} 
where $\gamma_d = \exp(2\pi i/d)$. Hence, it remains only to show that in high dimensions zero phase-error rate leads to perfect privacy.

\begin{figure}[htbp!]
\centering
\captionsetup{justification = raggedright}
\includegraphics[width=0.45\textwidth]{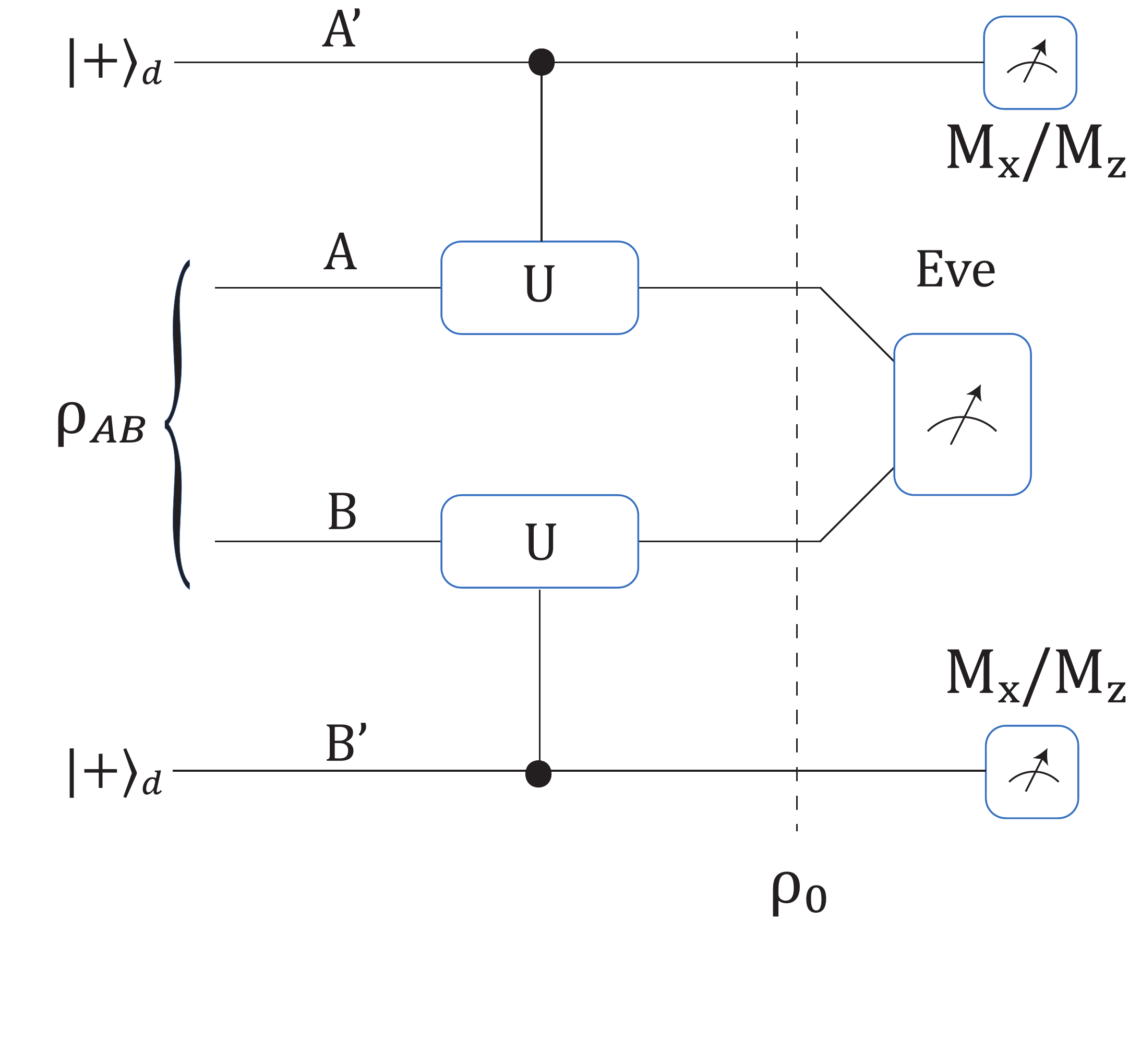}
\caption{Schematic diagram of the entanglement-based $d$-dimensional PM QKD, where $\rho_{AB}$ is a bipartite state on two optical modes, and the encoding operation $U$ rotates the coherent state by $2\pi/d$. The optical mode is phase rotated by $2\pi k/d$ if the $k$-th control dit is triggered. The encoded state $\rho_0$ is sent to the untrusted Eve for measurement, who is supposed to announce the key difference $(\kappa_a - \kappa_b) \text{ mod } d$, where $\kappa_a$ and $\kappa_b$ refer to the control dits triggered in $A'B'$. Alice and Bob distill secure keys from the qudit systems $A'$ and $B'$. The setup extends the binary entanglement-based symmetric encoding protocol in Ref.~\cite{Zeng2019Symmetryprotected}.}
\label{EBSymmetric}
\end{figure}

\subsection{Phase error and privacy in prime power dimensions}
\label{sec3B}

Phase error implies information leakage in two dimensions \cite{lo1999Unconditional, shor2000Simple, koashi2009simple}. In the security proof of two-dimensional QKD based on complementarity \cite{koashi2009simple}, the two-body entanglement distillation is first squashed into an equivalent single-body coherence distillation protocol given efficient bit-error correction \cite{ma2019operational}. As long as the squashed state is close to the two-dimensional $X$-basis eigenstate, they can share random and private keys after $Z$-basis measurements. This essentially connects privacy with phase error, i.e. the error in estimating the $X$ measurement results. Due to the anticommutability of the two-dimensional $X$ and $Z$ observables, the final $Z$ measurement anticommutes with the phase-error correction, and we can move the $Z$-basis measurement forward and reduce the phase-error correction to random hashing.

It is not obvious to generalize the two-dimensional complementarity argument to high dimensions. Efforts have been made in proving the security of prime-power-dimensional BB84 protocols \cite{Chau2005HDQKD,Nikolopoulos2005}, which implicitly connect privacy with phase error in prime power dimensions. Yet none of these early works have given an explicit distillable key-rate formula. In Appendix~\ref{Appendix2}, we give a simple justification of prime-power-dimensional phase-error correction based on the complementarity approach with parity check operations $\{P_l(\mathbf{v})\}_{l = 0}^{d-1}$,
\begin{equation}
P_l(\mathbf{v}) = \sum_{\mathbf{z}\cdot\mathbf{v} = l} \ket{\mathbf{z}}\bra{\mathbf{z}}.
\end{equation}

We thus yield a similar secure key-rate formula as two dimensions:
\begin{equation}
\label{key rate}
r = \log_2d- H(\vec{E}_{ph}) - H(\vec{E}_{bit}),
\end{equation}
where $H$ is the $\log_2$ based Shannon-entropy function. $\vec{E}_{ph}$ and $\vec{E}_{bit}$ are the phase and bit-error vector. They are defined as the length-$d$ error rate vectors of different shifts. To be specific, under an asymptotic setup, suppose Alice and Bob measure their $N$ pairs of qudit systems in $X$ basis and obtain two length-$N$ $d$-dimensional strings $\mathbf{a}$ and $\mathbf{b}$, the phase-error rate vector is defined as $\vec{E}_{\text{phase}} := \vec{wt}(\mathbf{a}-\mathbf{b})/N$, where the subtraction is of modulus $d$, and the weight function is defined as:
\begin{equation}
 \vec{wt}(\mathbf{s}) = 
 \begin{bmatrix}
 \text{No. of 0 in $\mathbf{s}$}\\
 \text{No. of 1 in $\mathbf{s}$}\\
 \cdots\\
  \text{No. of $d$-1 in $\mathbf{s}$}
 \end{bmatrix}
\end{equation}

The bit-error rate vector $\vec{E}_{bit}$ can be defined similarly. Note that a similar formula was derived in Ref. \cite{Sheridan2010Qudit} from information theoretic approaches.

We thus arrive at the following relation between phase error and privacy in prime power dimensions:
\begin{observation}
\label{Theorem2}
In prime power dimensions, if Alice can determine Bob's $X$-basis measurement results, i.e. there is no phase error, the protocol generates secure keys in $Z$ basis.
\end{observation}
This, combined with Observation~\ref{Theorem1}, proves the security of $d$-dimensional PM QKD. 

\subsection{Parameter estimation with decoy states}  
\label{sec3C}

In practice, the key-rate formula, Eq.~\eqref{key rate}, cannot be used directly as the phase-error vector $\vec{E}_{ph}$ based on the hypothetical qudit systems is not experimentally accessible. We can however estimate it based on the realistic optical mode systems. To be more specific, Alice and Bob can run discrete phase randomization where they independently add one of $D$ random phases $\phi_a$ and $\phi_b$ to their coherent states. This essentially transforms their states into mixtures of pseudo-Fock states (see Appendix~\ref{AppendixC2} and \ref{Appendix:discrete randomization}), which are symmetric states that yield no phase error as previously argued. With light intensity $\mu$, denote the fraction of detection caused by the $l$-photon state as $q_l^\mu$, which gives the length-$d$ vector $\vec{q}_{\mu}$ with $q_l^\mu$ at its $l$-th entry. We conclude that the phase-error vector $\vec{E}_{ph}$ is exactly the detection ratio of each joint Fock state $\vec{q}_{\mu}$. We thus have the experimentally accessible key-rate formula:

\begin{equation}
\label{final key rate}
r = \frac{d}{D}Q_\mu\{\gamma[\log_2d-H(\vec{E}^\mu_{bit})] -H(\vec{q}_{\mu})\},
\end{equation}
where the parameter $\gamma < 1$ marks the information reconciliation efficiency. The bit-error vector $\vec{E}^\mu_{bit}$ can be retrieved directly from random sampling. To access the detection fraction vector $\vec{q}_{\mu}$ of each symmetric state, we apply the decoy method \cite{Lo2005Decoy,wang2005decoy} by varying the light intensity $\mu$. This leads to the following high-dimensional PM QKD protocol with decoy states for parameter estimation:

\underline{\textbf{High-dimensional PM QKD protocol with}}

\underline{\textbf{parameter estimation}}
\begin{enumerate}
\item \textbf{Encoding}: Alice randomly generates a key ``dit" $\kappa_a$ from $\{0,1,\cdots,d-1\}$ and a random intensity $\mu_a$ as in the decoy method. She prepares the coherent state $\ket{\sqrt{\mu_a/2}~e^{i\frac{2\pi}{d} \kappa_a}}_A$. Similarly, Bob randomly picks $\kappa_b$ and $\mu_b$, and prepares $\ket{\sqrt{\mu_b/2}~e^{i\frac{2\pi}{d} \kappa_b}}_B$.
\item \textbf{Discrete phase randomization}: Alice and Bob independently phase randomize their coherent states for a large enough phase slice number $D$. That is, they randomly pick $\phi_a$ and $\phi_b$ from $\{j\frac{2\pi}{D}\}_{j=0}^{D-1}$ and prepare $\ket{\sqrt{\mu_a/2}~e^{i(\phi_a+\frac{2\pi}{d} \kappa_a)}}_A$ and $\ket{\sqrt{\mu_b/2}~e^{i(\phi_b + \frac{2\pi}{d}\kappa_b)}}_B$ respectively.
\item \textbf{Measurement}: Alice and Bob send the two optical modes $AB$ to an untrusted party, Eve, who is supposed to perform interference measurement and announce the detection results: no click, double click, $L$ click or $R$ click. 
\item \textbf{Sifting}: After many rounds of quantum communications, Alice and Bob keep only the rounds with $L$ or $R$ click. They announce the random intensities and phases $\mu_a$, $\phi_a$ and $\mu_b$, $\phi_b$ publicly. They keep only the rounds with $\mu_a = \mu_b$. For each intensity group, they postselect the rounds where $|\phi_a - \phi_b| \in \{k\frac{2\pi}{d}\}_{k=0}^{d-1}$. They end up with two correlated $d$-dimensional strings.
\item \textbf{Parameter estimation}: From the raw data they retained, Alice and Bob retrieve the gain $Q_\mu$ and the bit-error rate vector $\vec{E}^\mu_{bit}$. They estimate the phase-error rate vector $\vec{q}_\mu$ based on Eq.~\eqref{yield} and \eqref{fraction}. 
\item \textbf{Key generation}: Based on the parameter estimation results, Alice and Bob reconcile their raw strings by consuming certain secure keys. They then perform privacy amplification to extract the secure final keys from the reconciled keys.  
\end{enumerate}

Notice that after Eve's announcement of detection results, they announce the random phases and postselect the phase-matched rounds where $|\phi_a - \phi_b|\in \{k\frac{2\pi}{d}\}_{k=0}^{d-1}$. Since Eve announces only the detection results without access to the random phases, the overall phase-error rate does not depend on the later phase announcement, and so does the privacy \cite{Ma2012alternative,Ma2018phase,maeda2019repeaterless}. 

To estimate $q_l$, denote the yield of $l$-photon states as $Y_l$, the overall gain $Q_\mu$ can thus be expressed as:
\begin{equation}
\label{yield}
Q_\mu = \sum_{l=0}^{\infty} P_\mu(l)Y_l,
\end{equation}
where $P_\mu(l)$ denotes the source statistics of $l$-photon states. We can thus apply decoy methods by varying intensity $\mu$ to estimate the yield of each joint Fock state \cite{Lo2005Decoy,wang2005decoy}, and the fraction of detection is given by:
\begin{equation}
\label{fraction}
q^\mu_l = P_\mu(l) \frac{Y_l}{Q_\mu}.
\end{equation}  

Although it requires infinite decoy levels to estimate each $q_l$ exactly, since the optimal coherent light intensity is far below 1, three or more-photon components are negligible in the source, and hence in the detected signals. Therefore, three decoy levels are enough to estimate the phase-error vector $\vec{q}_{\mu}$ (see also the finite-size analysis in Ref.~\cite{Zeng2019Symmetryprotected}) and the detection fraction $\vec{q}_{\mu}$ is to be estimated with decoy states based on Eq.~\eqref{yield} and \eqref{fraction}. 

As the pseudo-Fock states given in Eq.~\eqref{pseudo Fock states} vary with the change in light intensity $\mu$, Eve may have chances to distinguish between signal and decoy states, thus cracking the decoy method \cite{Cao2015discrete}. Therefore, we want the generated pseudo-Fock states to be close to the real Fock states, i.e. we want the discrete randomization close to continuous. As shown in Appendix~\ref{Appendix:discrete randomization}, $D \geq 10$ is required for a negligible effect of discrete randomization. When $d = 2$, this essentially introduces a significant sifting factor $2/D$. For high-dimensional PM QKD that $d \geq 10$, however, we can simply let $D = d$ and the postselection can be omitted since $\phi_a$ and $\phi_b$ are themselves code phases. This manifests the simplicity in implementing high-dimensional PM QKD.  

\section{REFERENCE-FRAME INDEPENDENCE UNDER REALISTIC fiber SETUPS}
\label{4Comparison}
We demonstrate, with simulation, that without phase postcompensation, the high-dimensional PM QKD sufficiently achieves reference-frame independence \cite{Laing2010RFI}. We mainly consider two practical scenarios: fixed-phase misalignment and small phase fluctuation. The fixed-phase misalignment corresponds to the intrinsic reference system mismatch and the phase fluctuation is a random phase drift added by the fiber that is independent of the encoding, both assumed to be controlled by the adversary. By virtue of the encoding symmetry, the phase-error rate is decoupled with channel noise~\cite{Zeng2019Symmetryprotected}, that is, the bit-error patterns. Hence, phase misalignment affects only the bit-error rate, whilst the phase-error rate depends merely on light intensity. We show that fixed phase misalignment does not increase the bit-error rate of the high-dimensional PM QKD (Section~\ref{intrinsic}). Although phase fluctuation does add to its bit-error rate, the decrease in key rate is smaller than that of two-dimensional PM QKD due to the concavity of Shannon-entropy function (Section~\ref{fluctuation}).

To justify our arguments, we simulate the asymptotic performance of 17-dimensional PM QKD against two-dimensional without phase postcompensation. The simulation model is similar to that illustrated in Appendix~B of Ref.~\cite{Ma2018phase}, with parameters given in Table~\ref{Simpara}. A detailed description is placed in Appendix~\ref{App:SimFormula}. The key-rate formula generally follows Eq.~\eqref{final key rate}.

\subsection{IMMUNITY TO FIXED-PHASE MISALIGNMENT}
\label{intrinsic}
\begin{figure}[htbp!] 
\includegraphics[width=0.45\textwidth]{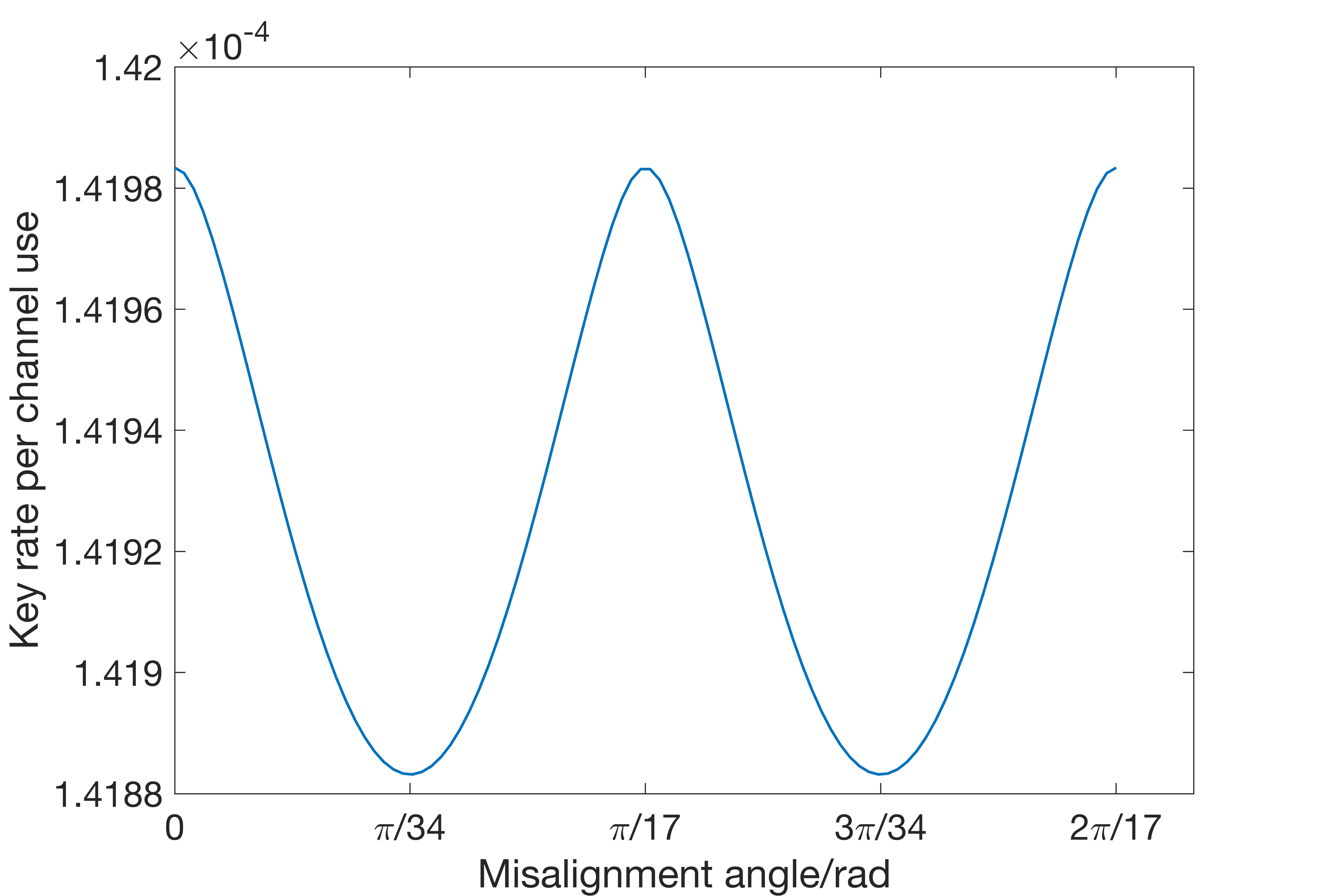}
\captionsetup{justification = raggedright}
\caption{Key rate of the 17-dimensional PM QKD at 100 km against fixed misalignment. The $\pi/17$ misalignment correlates the opposite key phases by $R$ clicks, and hence giving almost no effect on the key rate. The worst-case misalignment is reached at $\pi/34$, whose relative effect is negligible.}
\label{Misalign}
\end{figure}

We demonstrate that the high-dimensional PM QKD achieves almost completely immunity to fixed-phase misalignment, in clear contrast with the two-dimensional PM QKD, which is sensitive to phase-reference mismatch. In the two-dimensional PM QKD, the worst case scenario is that Alice and Bob hold phase references that differ by $\delta = \pi/2$. The protocol would not correlate Alice and Bob's keys. Suppose Alice sends phase $A_0$, it can be seen that no matter Bob sends $B_0$ or $B_1$, the interference result would highly likely be double clicks, and any single click does not provide too much information that helps Alice to distinguish Bob's key bit. However, in a $d$-dimensional PM QKD, suppose the phase references are differed by $\delta + \frac{2k\pi}{d}$ with $\delta \in [0,\frac{2k\pi}{d})$ and $k$ being integer. Note the $\frac{2k\pi}{d}$ term results only in a deterministic shift between key phases, and therefore can be tackled by classical postprocessing. Hence, the effective misalignment only ranges in $[0,\frac{2k\pi}{d})$, which gets smaller as $d$ increases, as shown in Fig.~\ref{Immunity} below. What is more, for the 17-dimensional PM QKD, we plot in Fig.~\ref{Misalign} the key rate at 100 km against misalignment ranging from 0 to $2\pi/17$. It can be seen that the lowest key rate is reached when the misalignment is $\pi/34$, which is one fourth between two key phases. This is reasonable since when the misalignment is half between two key phases at $\pi/17$, the $A_0$ phase would be determinedly matched to $B_8$ as they differ by $\pi$, causing $R$ click. Hence, $\pi/34$ is the worst-case misalignment right between the two deterministic misalignment 0 and $\pi/17$. It can be seen from Fig.~\ref{Misalign} that the effect of the fixed misalignment to the key rate of the 17-dimensional PM QKD is of 0.1\% scale, and hence negligible in practice.     

\begin{figure}[htbp!] 
\includegraphics[width=0.45\textwidth]{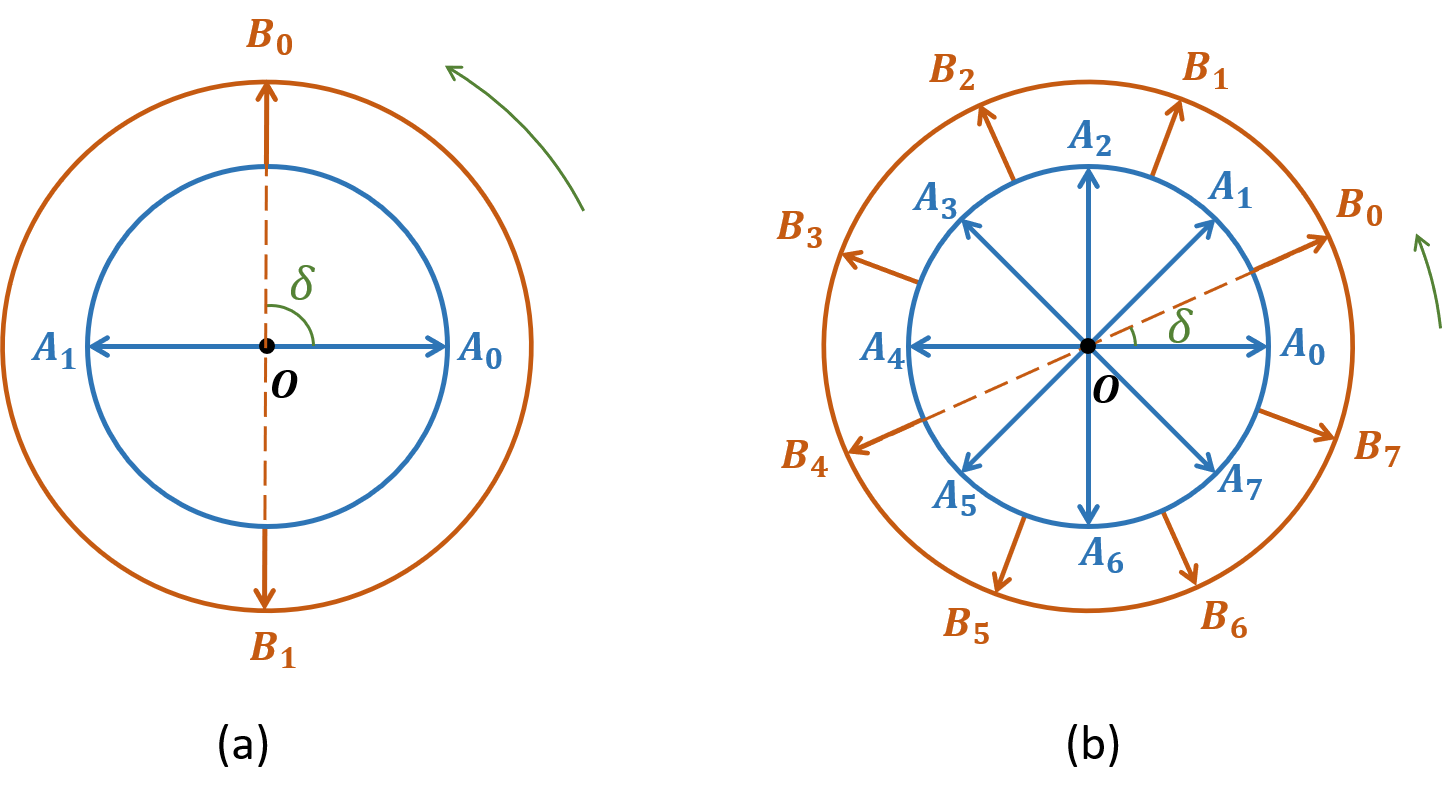}
\captionsetup{justification = raggedright}
\caption{Encoding circles of low- and high-dimensional PM QKD against worst-case misalignment. In the low-dimensional case, both encoding phases are far away from the deviated phase locations, thus giving much uncertainty. Yet in the high-dimensional case, the deviated phases are closer to key phases, enabling the error correction to coordinate the phase shift.}
\label{Immunity}
\end{figure}

To justify the above argument, we simulate the asymptotic performance of the two-dimensional PM QKD against the 17-dimensional PM QKD under various fixed misalignment compared with the linear repeaterless bounds~\cite{pirandola2017fundamental,takeoka2014fundamental}. The linear
bound we used here is the PLOB bound~\cite{pirandola2017fundamental},
which corresponds to the secret key capacity of the lossy channel. As shown in Fig.~\ref{Sim} below, without phase postcompensation, the key rate of two-dimensional PM QKD decreases continuously as the fixed misalignment increases. When the misalignment reaches $\pi/4$, the key rate of binary PM QKD generally discounts by a factor of 10, and when it further increases to $\pi/3$, the two-dimensional PM QKD cannot break the linear bound anymore. In clear contrast, the 17-dimensional PM QKD is almost completely immune to any phase misalignment. As can be seen in the figure, the 17-dimensional PM QKD performs almost identically under $\pi/34$ misalignment (the worst case) and no misalignment. Its key rate is similar to that of the perfectly aligned two-dimensional PM QKD, despite a slight decrease in the maximal reachable distance. On the other hand, the two-dimensional PM QKD clearly cannot generate any keys under $\pi/2$ misalignment. This demonstrates the superiority of high-dimensional PM QKD in terms of immunity to fixed misalignment.  

\begin{figure}[htbp!] 
\includegraphics[width=0.5\textwidth]{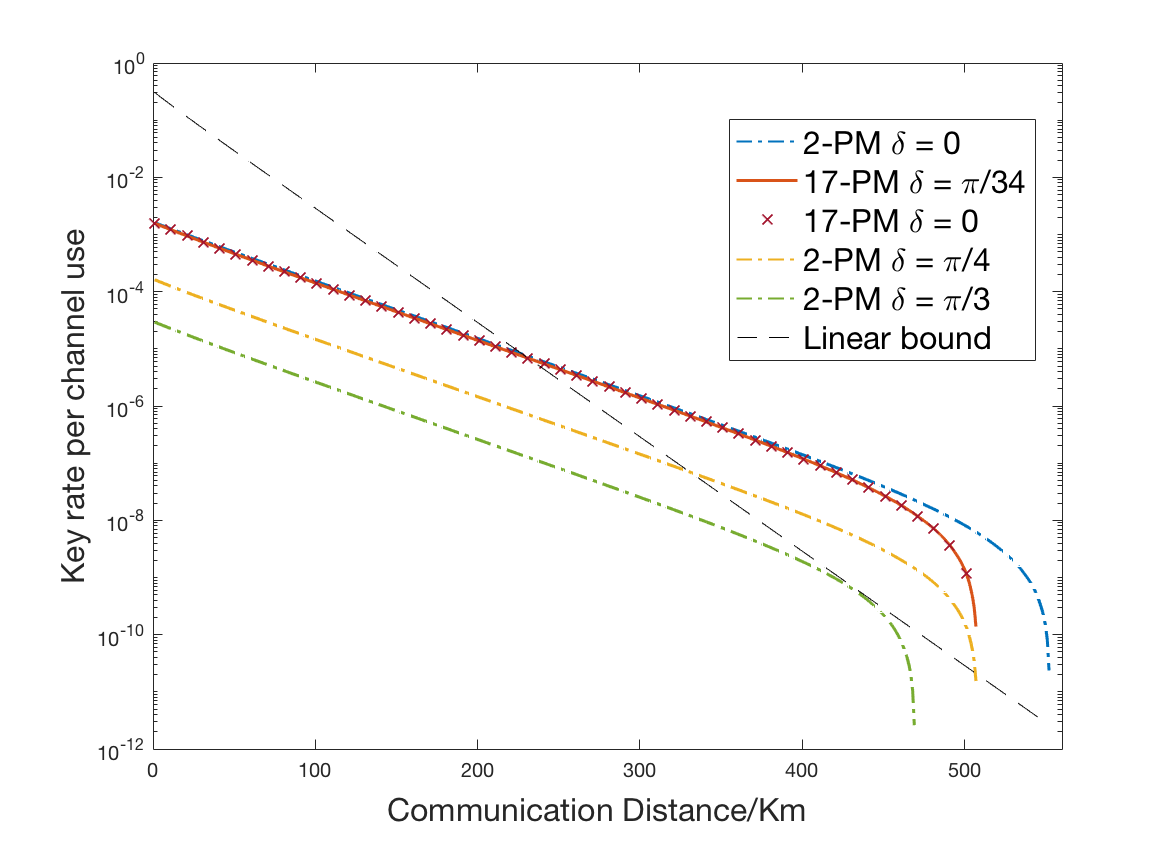}
\captionsetup{justification = raggedright}
\caption{Rate-distance performance of two- and 17-dimensional PM QKD against various fixed-phase misalignment, in comparison with the linear key-rate bounds \cite{takeoka2014fundamental,pirandola2017fundamental}. The linear bound we use in the plot is the PLOB bound~\cite{pirandola2017fundamental}. The key-rate performance of the 17-PM under the worst-case $\pi/34$ misalignment is similar to that of the 2-PM with no misalignment. The key rate of 2-PM decreases gradually and cannot generate keys at the worst-case $\pi/2$ misalignment.}
\label{Sim}
\end{figure}

\begin{table}[htbp!]
\caption{Summary of parameters used in the simulation}
\begin{tabular}{l c}
\hline 
{\bf Parameters} & {\bf Values} \\
\hline
fiber attenuation $\alpha$ & 0.2 dB/km\\
Dark count rate $p_d$ & $1\times10^{-8}$\\
Error correction efficiency $\gamma$ & 0.95\\
Detector efficiency $\eta_d$ & 20\%\\
No. of phase slices $D$ & 16\\
\hline 
\end{tabular}
\label{Simpara}
\end{table}

\subsection{Robustness to small phase fluctuation}
\label{fluctuation}
\begin{figure*}[htbp!] 

\begin{subfigure}{0.45\textwidth}
\includegraphics[width=\textwidth]{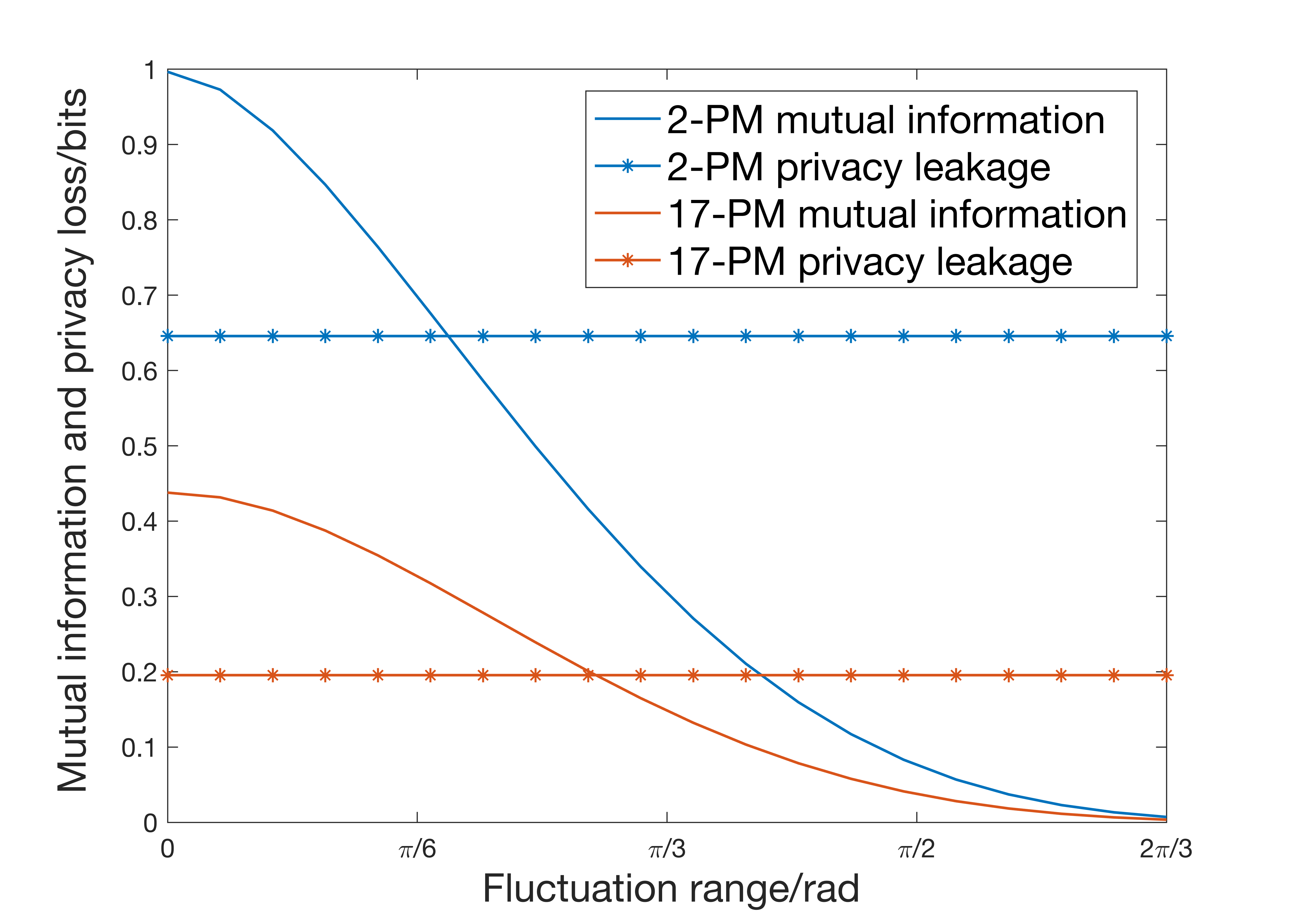}
\caption{}
\label{Fig:MIPER_fluc}
\end{subfigure}
\begin{subfigure}{0.45\textwidth}
\includegraphics[width=\textwidth]{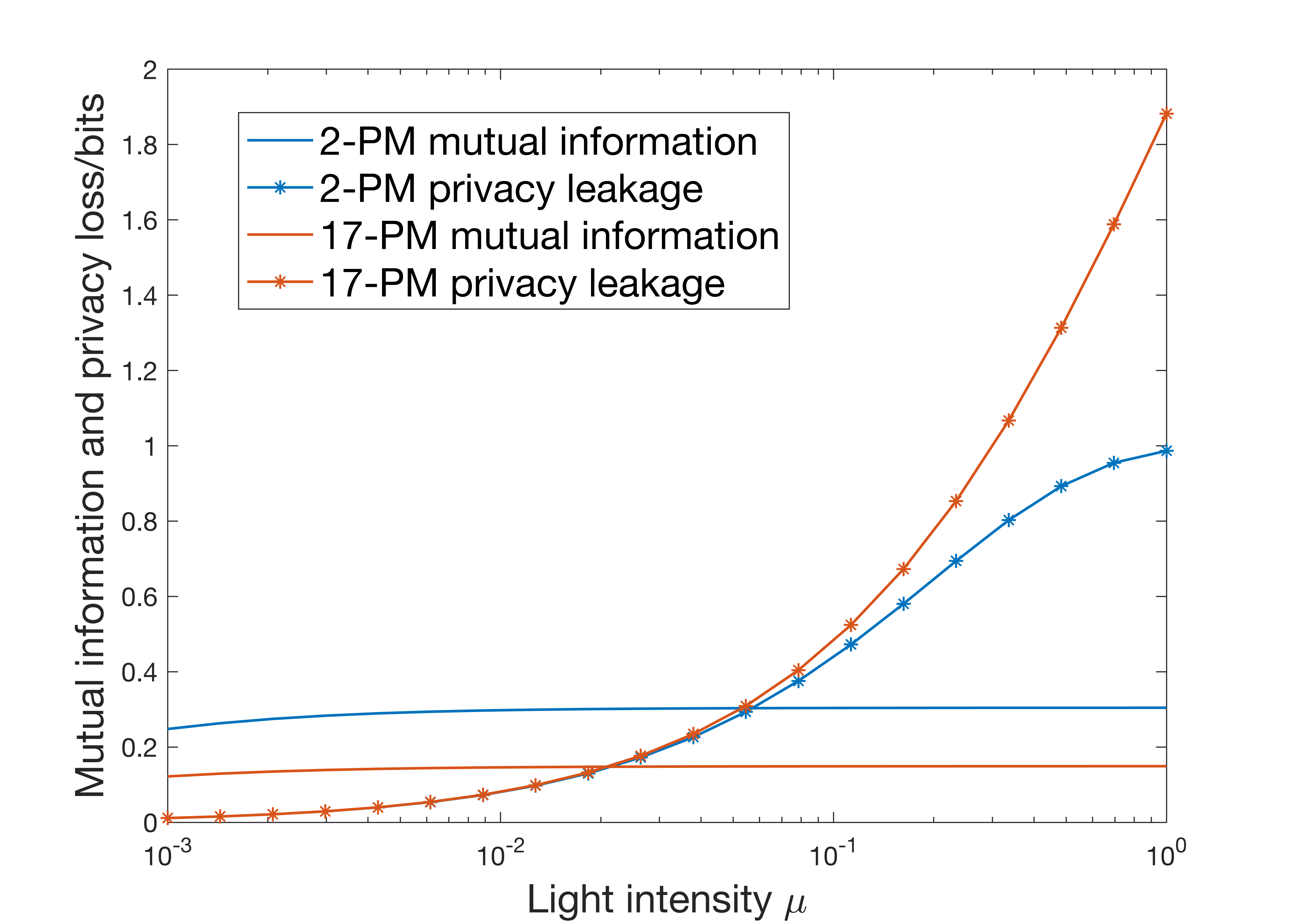}
\caption{}
\label{Fig:MIPER_mu}
\end{subfigure}

\begin{subfigure}{0.45\textwidth}
\includegraphics[width=\textwidth]{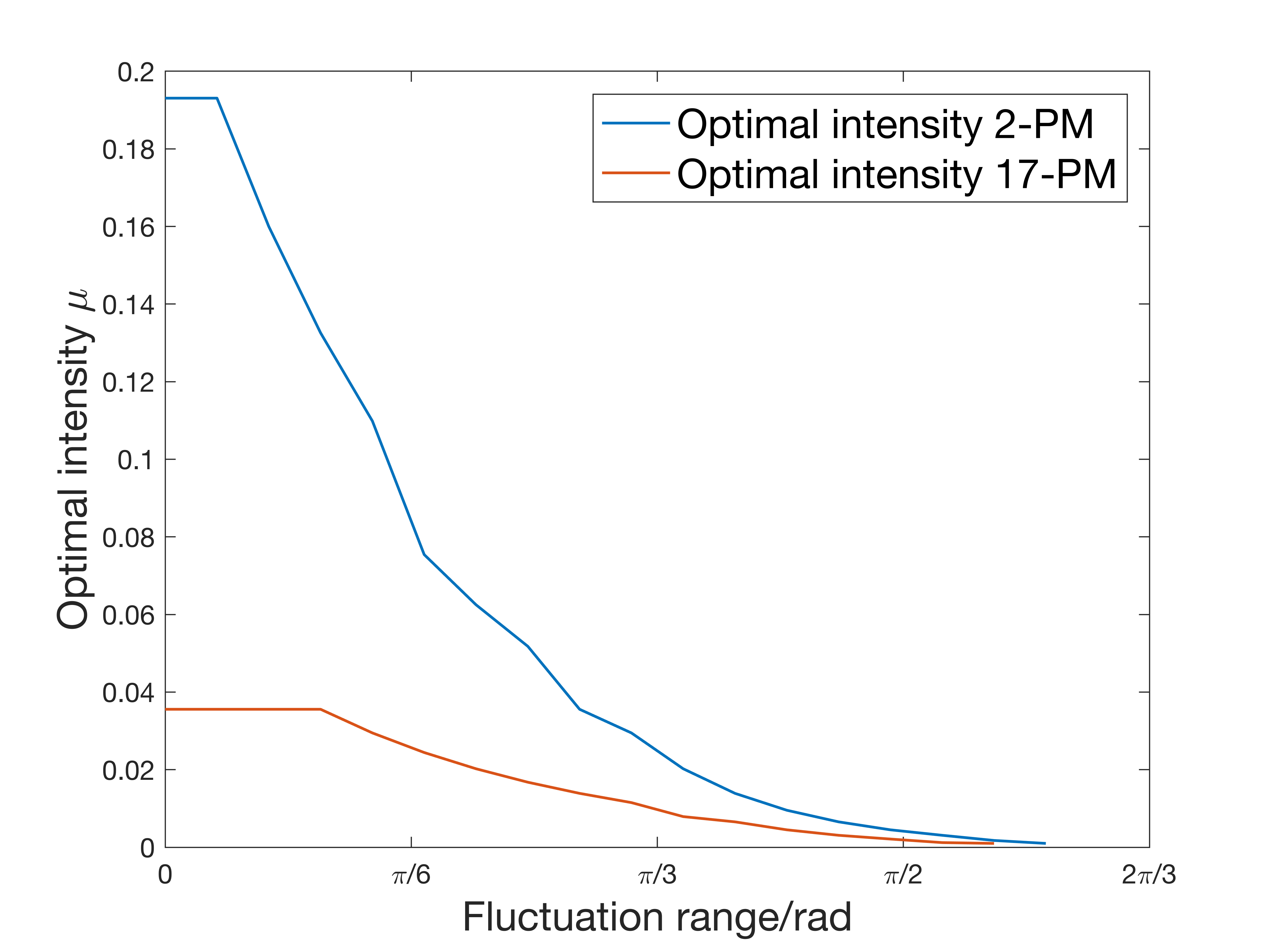}
\caption{}
\label{Fig:mus}
\end{subfigure}
\begin{subfigure}{0.45\textwidth}
\includegraphics[width=\textwidth]{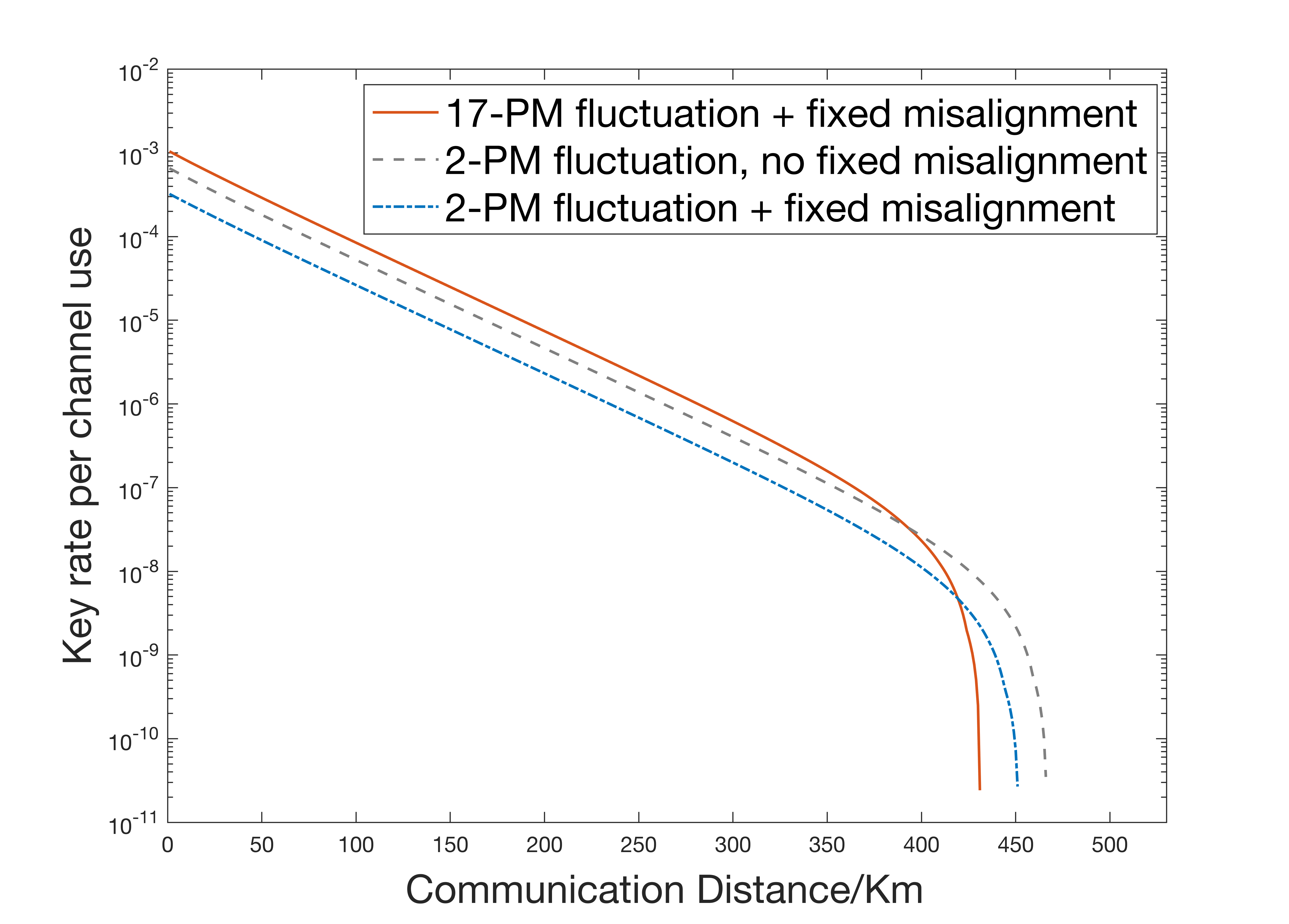}
\caption{}
\label{Fig:fluc_compare}
\end{subfigure}
\captionsetup{justification = raggedright}
\caption{High-dimensional PM QKD under phase fluctuation. \textbf{(a)} Mutual information and privacy leakage against phase fluctuation range $\phi_{lim}$ for 2- and 17-PM, light intensity 0.2 and 0.03 respectively, communication distance 300 km. The fluctuation does not affect the privacy as a result of encoding symmetry. The mutual information of the 2-PM decreases faster than that of the 17-PM. \textbf{(b)} Mutual information and privacy leakage against light intensity $\mu$ for 2- and 17-PM, fluctuation range $\pi/3$, communication distance 300 km. The light intensity does not affect the mutual information. \textbf{(c)} Optimal light intensity $\mu$ against phase fluctuation range $\phi_{lim}$ for 2- and 17-PM, communication distance 300 km. The 2-PM light intensity decreases rapidly as the fluctuation increases in order to compensate its faster drop in mutual information. \textbf{(d)} Simulated key-rate performance: 17-PM under both $\pi/6$ fixed misalignment and $\pi/3$ range fluctuation (red line), 2-PM under the same scenario (blue dotted line), 2-PM under fluctuation only, no fixed misalignment (gray dotted line). The 17-PM is superior than the 2-PM under small phase fluctuation.}
\label{Fig:Fluc}
\end{figure*}

When phase fluctuation is applied, the original code phases can no longer be recovered exactly since the fluctuation is random within a range of angles. In reality the phase fluctuation may come from the sources and the fiber, whilst the latter is length dependent. To illustrate our ideas, we adopt a simplified model that during each round a random noisy phase (independent of encoding) uniformly distributed in $[-\phi_{lim},\phi_{lim}]$ is added to the encoded coherent state, for Alice and Bob respectively. We focus on the term $\log_2(d) - H(\vec{E}_{bit})$, which denotes the mutual information between Alice and Bob, and the term $H(\vec{q_{\mu}})$, which denotes the cost due to phase-error rate, i.e., the leak of raw key information. Fixing the communication distance at 300 km, we compare in Fig.~\ref{Fig:MIPER_fluc} the change in mutual information $\log_2(d) - H(\vec{E}_{bit})$ and privacy leakage $H(\vec{q_{\mu}})$ for two- and 17-dimensional PM QKD against the phase fluctuation range $\phi_{lim}$. The light intensity $\mu$ is fixed at 0.2 and 0.03, respectively, for the two- and 17-dimensional, which is around the optimal value under no fluctuation shown in Fig.~\ref{Fig:mus}. It can be seen that the privacy leakage term $H(\vec{q_{\mu}})$ remains unchanged for both the two-dimensional and 17-dimensional regardless of the fluctuation range. This demonstrates the property of the encoding symmetry analysis (Section~\ref{sec3A}) that it decouples channel disturbance from privacy leakage \cite{Zeng2019Symmetryprotected}, and hence the fluctuation from the channel does not affect privacy. 

Notice that the two-dimensional has greater privacy leakage than the 17-dimensional. This is reasonable since in the two-dimensional key space the adversary is essentially guessing between two symbols, which is significantly easier than the 17-dimensional case where she guesses between 17 symbols. In contrast, the mutual information term $\log_2(d) - H(\vec{E}_{bit})$ drops for both cases, as the fluctuation clearly results in higher bit error. We see that the mutual information of the two-dimensional is higher than that of the 17-dimensional, which implies that the two-dimensional has fewer bit errors. This can be understood as the single-photon interference detector outputs binary results (left or right click), and thus it is ideal for binary key space and yields very low bit-error rate for the two-dimensional protocol when no fluctuation is applied (the mutual information is close to 1 bit as shown in the figure). It however does not provide full information for the 17-dimensional protocol unless the input coherent states are in the same or opposite phases. It thus generates lower mutual information for the 17-dimensional than the two-dimensional, although their overall key rates are similar since the 17-dimensional has lower privacy leakage. Moreover, the mutual information of the two-dimensional PM QKD decreases more rapidly than that of the 17-dimensional. This is reasonable since the bit-error rate of the two-dimensional is very low under no fluctuation. Yet when fluctuation adds to its bit-error rate, the change rate in the term $H(\vec{E}_{bit})$ is significantly higher since the derivative of the Shannon-entropy function $H(p)$ is infinity when $p$ tends to 0. Hence, we see in Fig.~\ref{Fig:MIPER_fluc} that the mutual information of two-dimensional PM QKD drops more rapidly than that of the 17-dimensional. 

In order to cope with the drop in mutual information, the privacy leakage term has to be lowered, which can be achieved through suppressing the intensity $\mu$ of the source. Fig.~\ref{Fig:MIPER_mu} illustrates the effects of light intensity $\mu$ on the mutual information and privacy leakage. The channel distance is fixed at 300 km, and a phase fluctuation of range $\phi_{lim} = \pi/3$ is applied. As expected from the encoding symmetry analysis, the mutual information term generally does not relate with the light intensity. As the light intensity drops, the single-photon fraction from the light source increases, and so does the single-photon fraction in the detection. This further lowers the uncertainty in the detection fraction of each photon number state $\vec{q}_\mu$, i.e. it lowers the privacy leakage $H(\vec{q_{\mu}})$, as shown in Fig.~\ref{Fig:MIPER_mu}. In order to compensate the faster drop in mutual information of two-dimensional PM QKD, its source intensity has to decrease further than that of the 17-dimensional, as shown in Fig.~\ref{Fig:mus}. 

The drop in the intensities results in a further drop in the overall gain $Q_{\mu} \approx \eta\mu$. Hence, as shown in Fig.~\ref{Fig:fluc_compare}, under a small phase fluctuation of range $\phi_{lim} = \pi/3$, the 17-dimensional PM QKD yields higher secure key rates than the two-dimensional. Moreover, when fixed misalignment is introduced, the key rate of two-dimensional PM QKD decreases further, whilst that of the 17-dimensional remains. We thus conclude that the high-dimensional PM QKD is more robust to small phase fluctuation than the two-dimensional PM QKD.  

\section{Concluding remarks}
\label{5Conclusion}
We generalize the two-dimensional PM QKD to any prime dimensions and analyze their asymptotic performance. Under a reasonable fiber-optic simulation setup, we demonstrate that when the protocol dimension is high enough, the key-rate performance is almost completely immune to fixed phase-reference-mismatch and robust to small phase fluctuation, i.e. it is reference-frame-independent. In general, our work points out the feasibility of increasing protocol dimension in order to combat misalignment. Our security argument provides the theoretical cornerstone for the analysis of high-dimensional QKD protocols. As possible extensions of this work, we discuss the following two remarks. 

Although in the general security proof we cover all the systems of prime power dimensions, we analyze only prime-dimensional PM QKD. This is due to the incompatibility of the rotating encoding and the additive group of prime power finite fields. For instance, the encoding operations of a four-dimensional PM QKD form the order-4 cyclic group $\{I,U,U^2,U^3\} \equiv Z_4$, where $U$ is the $\pi/2-$rotation operator. In contrast, the additive group of GF(4) is the Klein-4 group $\{a,b|a^2=b^2=1\}$. This incompatibility invalidates Observation \ref{Theorem1} for prime power dimensions. One possible solution is to alter the encoding operations. For instance for four dimensions, the encoding operations can be changed to $\{I,U,V,UV\}$, where $U$ is the $\pi-$rotation operator, and $V$ satisfies:
\begin{equation}
V\ket{x+ip} = \ket{p+ix}
\end{equation}
Clearly, this encoding operation set is also valid. Since $U^2 = V^2 = I$, the encoding operation group is isomorphic to the Klein-4 group, and hence compatible with the addition in GF(4). In fact, it can be verified that Observation \ref{Theorem1} holds under this encoding set. However, the caveat is that the operation $V$, which essentially changes the two quadratures, is not unitary, and hence arguably not physical. We thus do not include the ambiguous prime power case in our security proof.     

The phase-encoding protocols such as PM QKD bear similarity as the discrete modulated MDI continuous-variable QKD (DM MDI CV QKD): replacing the interference detector by the CV-Bell-like detector \cite{Pirandola2015MDICV}, we obtain the DM MDI CV QKD introduced in Ref. \cite{Ma2019DMMDICV}. Since in the MDI setup, the detector does not affect the security, we can apply the security analysis in this work directly to DM MDI CV QKD. This opens up the prospective to prove CV-QKD security using DV methods, which is recently discussed in Ref. \cite{matsuura2020finitesize}.  

\begin{acknowledgments}
A.J. and R.V.P. acknowledge support from the UK EPSRC Quantum 
Communications Hub, project EP/T001011/1. A.J. acknowledges funding from Cambridge Trust. P.Z. and X. M. acknowledge funding from the National Natural Science Foundation of China Grants No.~11875173 and No.~1217040781, the National Key Research and Development Program of China Grant No.~2019QY0702 and No.~2017YFA0303903.
\end{acknowledgments}

\begin{appendix}
\section{Definitions and mathematical backgrounds}
This section provides the essential mathematical tools, conventions and specific techniques employed in the security proof of high-dimensional QKD.
\subsection{Finite fields}
\label{AppendixA1}
The finite field, or Galois field, is the algebraic structure that lies in the discrete-value information processing. In a general $d$ dimensional information processing task, the set $\{0,1,\cdots,d-1\}$ are the symbols. In order to construct an algebra on this set, we need to define properly addition $\oplus$ and multiplication $\odot$ operations such that they follow the usual associative, commutative and distributive laws and each has identity and inverse. In other words, we need to make the symbol set a finite field, denoted by GF($d$), by defining the addition and multiplication operations. 

For prime dimension $p$, the set $\{0,1,\cdots,p-1\}$ can be made a finite field trivially equipped with the usual modulus $p$ addition and multiplication. This is the finite field $\mathbf{Z}_p$, and it can be seen that every GF($p$) is isomorphic to $\mathbf{Z}_p$.

Next, consider the prime power dimension $d = p^r$. We define the canonical addition on the set $\{0,1,\cdots,d-1\}$ such that:
\begin{equation}
\begin{aligned}
&a = \sum_{m = 0}^{r-1} a_m p^m \quad b = \sum_{m = 0}^{r-1} b_m p^m\\
& a\oplus b = \sum_{m = 0}^{r-1} (a_m\oplus_p b_m) p^m,
\end{aligned}
\end{equation}
where $\oplus_p$ is the $p$-modulus addition and $a_m, b_m$ are the $p$-ary decompositions of $a$ and $b$. This is a valid field addition for GF($p^r$). In fact, the field multiplication can also be constructed for GF($p^r$), and it can be shown that the set $\{0,1,\cdots,d-1\}$ can be made a field if and only if $d = p^r$, i.e only prime power degree finite fields exist \cite{artin2011algebra}.

The convenience of adopting the canonical addition defined above is its compatibility with exponential operations. We will encounter frequently the complex exponential $\gamma_p^a$, where $\gamma_p$ is the complex number such that $\gamma_p^p = 1$ and $a \in $GF($d$). The value of $\gamma_p^a$ is a complex number calculated as if $a$ were the usual integer. Note that the exponential multiplication rule follows:
\begin{equation}
\gamma_p^a \gamma_p^b = \gamma_p^{a+b} = \gamma_p^{a\oplus b},
\end{equation}
where $+$ is the integer addition and $\oplus$ is the canonical field addition. It can also be seen that the canonical field addition is also compatible with conjugation and distributive law in the way that:
\begin{equation}
\begin{aligned}
&(\gamma_p^a)^* = \gamma_p^{(-a)} = \gamma_p^{(\ominus a)}\\
&\gamma_p^{a\odot c}\gamma_p^{b\odot c} = \gamma_p^{a\odot c + b\odot c} = \gamma_p^{a\odot c \oplus b\odot c} = \gamma_p^{(a\oplus b)\odot c}
\end{aligned}
\end{equation}

Since we are always working with the complex exponential $\gamma_p^a$ in the security proof next section, we will use $+$ in replace of $\oplus$ as they are equivalent. The field multiplication is not compatible with complex exponential in the sense that $(\gamma_p^a)^b \neq \gamma_p^{a\odot b}$ (except for $\mathbf{Z}_p$). However, in the following discussions, we do not need operations like $(\gamma_p^a)^b$, and hence we will still replace $a\odot b$ as $ab$. 

\subsection{The Heisenberg-Weyl group: high-dimensional Pauli operators}
\label{AppendixA2}
We introduce the Heisenberg-Weyl group as a generalization of the two-dimensional Pauli group \cite{Durt2010MUB}. For a prime-power-dimensional space, i.e. $d = p^r$, with computational basis $\{\ket{l}\}_{l=0}^{d-1}$, we define
\begin{equation}
\begin{aligned}
Z &= \sum_{l=0}^{d-1} \gamma_p^{l} \ket{l}\bra{l}, \\
X &= \sum_{l=0}^{d-1} \ket{l+1}\bra{l}, \text{ with respect to GF($d$).}\\
\end{aligned}
\end{equation}

A natural mutually-unbiased basis (MUB) of $Z$-basis is given by the eigenbasis of $X$,
\begin{equation}
\begin{aligned}
\ket{\tilde{l}} :=& \frac{1}{\sqrt{d}} \sum_{j=0}^{d-1} \gamma_p^{-lj} \ket{j}, \\
\ket{j} =& \frac{1}{\sqrt{d}} \sum_{j=0}^{d-1} \gamma_p^{lj} \ket{\tilde{l}}. \\
\end{aligned}
\end{equation}
Note that $X\ket{\tilde{l}} = \gamma_p^l \ket{\tilde{l}}$. This is the basis complementary to the computational basis.

The Heisenberg-Weyl operator $W(u,v)$ is defined to be
\begin{equation}
W(u,v) = \sum_{l = 0}^{d-1} \ket{l + u}\gamma_p^{lv}\bra{l},
\end{equation}
with $u,v=0,1,...,d-1$. It is easy to verify that
\begin{equation}
W(u, 0)W(0,v) = \gamma_p^{-uv}W(0,v)W(u,0)
\end{equation}
In prime dimension this reduces to the usual identity:
\begin{equation}
X^uZ^v = \gamma_p^{-uv}Z^vX^u
\end{equation}

From the Heisenberg-Weyl operators, we can generate the Bell basis in prime power dimensions. Define $\ket{\Phi_{0,0}} = \ket{\Phi_+} = \dfrac{1}{\sqrt{d}}\sum_{j=0}^{d-1} \ket{jj}$. Write $\Phi_{0,0}$ in $X$-basis,
\begin{equation}
\begin{aligned}
\ket{\Phi_{0,0}} &= \frac{1}{\sqrt{d}} \sum_{j=0}^{d-1} \ket{jj} \\
&= \frac{1}{\sqrt{d}} \sum_{k,l=0}^{d-1} \sum_{j=0}^{d-1} \gamma_p^{j(k+l)} \ket{\tilde{k}\tilde{l}} \\
&= \frac{1}{\sqrt{d}} \sum_{k=0}^{d-1} \ket{\tilde{k},\widetilde{-k}}. \\
\end{aligned}
\end{equation}

The generalized qudit Bell states \cite{bennett1993Teleporting} are
\begin{equation}
\begin{aligned}
\ket{\Phi_{u,v}} &:= (I\otimes W(u,v))\ket{\Phi_+} \\
&= \frac{1}{\sqrt{d}}\sum_{l=0}^{d-1} \gamma_p^{lv}\ket{l}_A\otimes \ket{l+u}_B,
\end{aligned}
\end{equation}
Denote $\Phi_{u,v}:=\ket{\Phi_{u,v}}\bra{\Phi_{u,v}}$. The qudit Bell states $\{\Phi_{u,v}\}_{u,v=0}^{d-1}$ form an orthonormal basis,
\begin{equation}
\begin{aligned}
\braket{\Phi_{u,v}|\Phi_{u',v'}} &= \frac{1}{d} \sum_{m,l = 0}^{d-1} \gamma_p^{-lv}\gamma_p^{mv'}\bra{l,l+u}\ket{m,m+u'}\\
&= \frac{1}{d}\sum_{l=0}^{d-1} \gamma_p^{-l(v-v')}\delta_{u_d,0}\\
&= \delta_{u_d,0}\delta_{v_d,0} 
\end{aligned}
\end{equation}
where $u_d:= u'-u, v_d:=v'-v$.

\subsection{Parity check in GF(d)}
\label{AppendixA3}
We discuss the parity check operations for qudits since it plays a central role in the security proof. A length-$N$ GF($d$) string $\mathbf{x}$ is an ordered $N$-tuple:
\begin{equation}
\mathbf{x} = [x_0,x_1,\cdots,x_{N-1}],
\end{equation}   
where each element belongs to GF($d$). For two length-$N$ GF($d$) strings $\mathbf{x}$ and $\mathbf{y}$, define the dot product as
\begin{equation}
\mathbf{x}\cdot\mathbf{y} = \sum_{k=0}^{N-1} x_k y_k,
\end{equation} 
where the additions and multiplications are defined on GF($d$).

We focus on non-zero parity check as zero parity check would give a zero result for sure. For a fixed non-zero $\mathbf{y}$, the dot product $\mathbf{x}\cdot\mathbf{y}$ gives $d$ different results uniformly, i.e. there are $d^{N-1}$ string $\mathbf{x}$ giving the same $\mathbf{x}\cdot\mathbf{y}$. We call this dot product the parity check of $\mathbf{x}$, and it can be seen that one praity-check equation divides the overall string space into $d$ cosets, each represented by the dot product result, which is a member of GF($d$).

According to linear algebra, in order to completely determine an unknown length-$N$ GF($d$) string $\mathbf{x}$, it takes $N$ linearly independent praity-check equations. This idea can be extended to quantum systems. Define the $Z$-parity measurement channel as the Kraus representation:
\begin{equation}
\mathcal{M}_{Z}(\mathbf{v})\{\rho\} = \sum_{l = 0}^{d-1} P_l(\mathbf{v})\rho P_l(\mathbf{v})^\dagger,
\end{equation} 
where $\mathbf{v}$ is a length-$N$ GF($d$) string and $\rho$ is any density operator on $\mathcal{H}_d^{\otimes N}$. The Kraus operator $P_l(\mathbf{v})$ is given as the projector onto the space of parity check result $\mathbf{z}\cdot\mathbf{v} = l$:
\begin{equation}
P_l(\mathbf{v}) = \sum_{\mathbf{z}\cdot\mathbf{v} = l} \ket{\mathbf{z}}\bra{\mathbf{z}}
\end{equation}

Similarly, we can define the $X$-parity measurement $\mathcal{M}_{X}(\mathbf{v})$ with respect to the $X$ basis. The output of the parity measurement is a mixture of $d$ parity states, i.e. 
\begin{equation}
\begin{aligned}
\mathcal{M}_{Z}(\mathbf{v})\{\rho\} &= \sum_{l = 0}^{d-1} P_l(\mathbf{v})\rho P_l(\mathbf{v})^\dagger\\
&= \sum_{l = 0}^{d-1} \sum_{\mathbf{z},\mathbf{z'}\cdot \mathbf{v} = l} \ket{\mathbf{z}}\bra{\mathbf{z}}\rho\ket{\mathbf{z'}}\bra{\mathbf{z'}}\\
&= \sum_{l = 0}^{d-1} p_l \rho_l,
\end{aligned}
\end{equation} 
where
\begin{equation}
\begin{aligned}
&p_l = \sum_{\mathbf{z}\cdot\mathbf{v}=l}\bra{\mathbf{z}}\rho\ket{\mathbf{z}}\\
&\rho_l = \sum_{\mathbf{z},\mathbf{z'}\cdot \mathbf{v} = l} \frac{\bra{\mathbf{z}}\rho\ket{\mathbf{z'}}}{p_l} \ket{\mathbf{z}}\bra{\mathbf{z'}} \text{, having parity $l$.}
\end{aligned}
\end{equation} 

It can be seen that it takes $N$ linearly independent $Z$/$X$-parity measurements to determine the $Z$/$X$ measurement results of an unknown state in $\mathcal{H}_d^{\otimes N}$. 

\section{Security proof of high-dimensional QKD}
\label{Appendix2}
In this section, we provide the rigorous security analysis of high-dimensional QKD. Our proof follows the outline of Koashi's complementarity approach \cite{koashi2009simple}, and extends it by generalizing the phase-error correction procedure to high dimensions.

\subsection{The squashing protocol}
The core of Koashi's qubit-based security proof is to reduce the two-body private and random key distribution to a single-body private and random number generation, i.e. to reduce entanglement distillation to coherence distillation \cite{ma2019operational}. The security of the actual protocol can thus be proved if the single body squashing protocol is secure. 

Consider the entanglement-based actual protocol below. Note that its equivalence to the prepare-and-measure protocol follows from the usual Shor-Preskill arguments \cite{shor2000Simple}. Also note that this protocol is of prime power dimension $d = p^r$ rather than qubit-based (we use ``digits" in replace of ``bits").  

{\bf\underline{Actual protocol}}
\begin{enumerate}
\item \textbf{State distribution} Alice and Bob share a bipartite state $\rho_{AB}$ in the space $(\mathcal{H}_A \otimes \mathcal{H}_B)^{\otimes N}$ after $N$ runs of quantum communications.
\item \textbf{Measurement} Alice and Bob measure their systems $\mathcal{H}_A^{\otimes N}$ and $\mathcal{H}_B^{\otimes N}$ respectively. They obtain two $N$-digit unreconciled key strings.
\item \textbf{Error correction} They reconcile the key strings through an encrypted classical channel consuming $l_{ec}$ digits of secret key. They agree on an $N$-digit raw key string $\kappa_{rec}$ except for a small failure probability $\epsilon_{cor}$.
\item \textbf{Privacy amplification} Alice randomly chooses $(N-m)$ $N$-digit strings $\{V_k\}_{k=1,\cdots,N-m}$, which are linearly independent, and announces them to Bob. The final key length is $(N-m)$, where the $k$-th key digit is $\kappa_{rec} \cdot V_k$, where the dot product is to be understood with addition and multiplication in the finite field $GF(d)$. Denote the final key as $\kappa_{fin}$.
\end{enumerate}

After the protocol, the overall state shared by Alice and Bob and Eve is
\begin{widetext}
\begin{equation}
\label{actualfin}
\rho_{ABKE}^{fin} = \sum_{\kappa_A, \kappa_B, \kappa_{fin}} Pr_{A, B, K}(\kappa_A, \kappa_B, \kappa_{fin})\ket{\kappa_A}_A\bra{\kappa_A}\otimes\ket{\kappa_B}_B\bra{\kappa_B}\otimes\ket{\kappa_{fin}}_K\bra{\kappa_{fin}}\otimes\rho_E(\kappa_A, \kappa_B, \kappa_{fin}),
\end{equation}
\end{widetext}
where $K$ is the key generation system (it is taken as Alice's system $A$ usually), and $Pr_{A, B, K}(\kappa_A, \kappa_B, \kappa_{rec})$ is the probability of Alice and Bob holding an $(N-m)$-digit string $\kappa_A$ and $\kappa_B$ respectively after the protocol whilst the correct reconciled key string being $\kappa_{rec}$. On the other hand, the ideal state is
\begin{equation}
\rho_{ABKE}^{ideal} = (d^{N-m})^{-1}\sum_{\kappa} \ket{\kappa}_A\bra{\kappa}\otimes\ket{\kappa}_B\bra{\kappa}\otimes\ket{\kappa}_K\bra{\kappa}\otimes\rho_E,
\end{equation}
where Alice and Bob share the correct reconciled key string, which is completely random and decoupled from Eve's system. 

In this way, a QKD protocol is defined to be $\epsilon$-secure, if the final distilled state $\rho_{ABKE}^{fin}$ is close to the ideal state $\rho_{ABKE}^{ideal}$ for a properly chosen $\rho_E$
\begin{equation}
\min_{\rho_E} \frac{1}{2}||\rho_{ABKE}^{fin} - \rho_{ABKE}^{ideal}||_1 \leq \epsilon,
\end{equation}
where $||A||_1 \equiv Tr\{\sqrt{A^\dagger A}\}$ is the trace norm.

In the error correction step of the actual protocol, we claim that Alice and Bob can correct their strings to $\kappa_{rec}$ except for a small failure probability $\epsilon_{cor}$. This means the protocol is $\epsilon_{cor}$-correct since
\begin{equation}
Pr_{A,B,K}(\kappa_A \text{ or }\kappa_B \neq \kappa_{rec}) \leq \epsilon_{cor}
\end{equation}
This property simply states Alice and Bob would very likely be sharing the same correct key strings. Hence intuitively, we can think Alice and Bob and the reconciled key generation system $K$ as a single party, i.e. we squash them into one system.  

More precisely, if Alice and Bob can apply a squashing operation $\Lambda$ on $(\mathcal{H}_A \otimes \mathcal{H}_B)^{\otimes N}$ to convert it to a key space $\mathcal{K}^{\otimes N}$ and an ancillary space $\mathcal{H}_R$, and the key measurement statistics on $\mathcal{K}^{\otimes N}$ is the same as $\kappa_{rec}$ in the actual protocol, we arrive at the following squashing protocol

{\bf\underline{Squashing protocol}}
\begin{enumerate}
\item \textbf{State distribution} Alice and Bob share a bipartite state $\rho_{AB}$ in the space $(\mathcal{H}_A \otimes \mathcal{H}_B)^{\otimes N}$ after $N$ runs of quantum communications.
\item \textbf{Squashing} They apply $\Lambda$ on $\rho_{AB}$ and convert it to a key space $\mathcal{K}^{\otimes N}$ and an ancillary space $\mathcal{H}_R$, i.e. $\Lambda(\rho_{AB}) \in \mathcal{K}^{\otimes N} \otimes \mathcal{H}_R$.
\item \textbf{Measurement} They measure $\mathcal{H}_R$ by $\mathcal{M}_R$ to obtain result $\gamma$. They then measure $\mathcal{K}^{\otimes N}$ to obtain $\kappa_{rec}$, with the same measurement statistics as that in the actual protocol. 
\item \textbf{Privacy amplification} They randomly choose $(N-m)$ $N$-digit strings $\{V_k\}_{k=1,\cdots,N-m}$, which are linearly independent. The final key length is $(N-m)$, where the $k$-th key digit is $\kappa_{rec} \cdot V_k$. Denote the final key as $\kappa_{fin}$.
\end{enumerate}
Since the key space $\mathcal{K}^{\otimes N}$ measurement statistics is the same as that of the actual protocol, the final state after the squashing protocol is therefore
\begin{equation}
\rho_{KE}^{fin} = \sum_{\kappa_{fin}} Pr_{K}(\kappa_{fin})\ket{\kappa_{fin}}_K\bra{\kappa_{fin}}\otimes\rho_E(\kappa_{fin}),
\end{equation}
where the probability distribution $Pr_{K}(\kappa_{fin})$ is the marginal distribution of $Pr_{A, B, K}(\kappa_A, \kappa_B, \kappa_{fin})$ in the actual final state \eqref{actualfin}, whilst the ideal squashed state is 
\begin{equation}
\rho_{KE}^{ideal} = (d^{N-m})^{-1}\sum_{\kappa} \ket{\kappa}_K\bra{\kappa}\otimes\rho_E,
\end{equation}
Likewise, the squashing protocol is termed $\epsilon_{sec}$-secret if the squashed state $\rho_{KE}$ is close to ideality, i.e.
\begin{equation}
\min_{\rho_E} \frac{1}{2}||\rho_{KE}^{fin} - \rho_{KE}^{ideal}||_1 \leq \epsilon_{sec}
\end{equation}
 
In \cite{koashi2009simple}, it is proved that as long as the squashing protocol is $\epsilon_{sec}$-secret with an $\epsilon_{cor}$-correct error correction, the actual protocol is $(\epsilon_{sec} + \epsilon_{cor})$-secure. Notice that in Ref. \cite{koashi2009simple} the quantum system is of dimension 2, but it can be trivially generalized to arbitrary dimension.

\subsection{The phase-error correction protocol}
It now remains to show that the single-body squashing protocol is secure. We do this by invoking phase-error correction, which bears intuitions from the uncertainty principle of two complementary operators: if the $X$-basis measurement of $\mathcal{K}^{\otimes N}$ is completely certain, the $Z$-basis measurement of it, which is by convention the key generation measurement, is completely random.

To be more specific, suppose $Z$-basis measurement on $\mathcal{K}^{\otimes N}$ is used for key generation in the squashing protocol. If, before the key generation measurement on $\mathcal{K}^{\otimes N}$, Alice and Bob are able to determine the $X$-basis measurement result of $\mathcal{K}^{\otimes N}$ to be $\mathbf{x}^*$ except for a small failure probability $\epsilon_T'$, we would have
\begin{equation}
\bra{\tilde{\mathbf{x}^*}} \rho_K \ket{\tilde{\mathbf{x}^*}} = F(\rho_K, \ket{\tilde{\mathbf{x}^*}} \bra{\tilde{\mathbf{x}^*}}) \geq 1 - \epsilon_T',
\end{equation}
i.e. the state on $\mathcal{K}^{\otimes N}$ is close to the $X$ eigenstate $\ket{\tilde{\mathbf{x}^*}}$ in terms of fidelity $F$. Following Ref. \cite{Fung2010Practical}, it can be shown that there exists $\sigma_E$ on Eve's system such that 
\begin{equation}
F(\rho_{KE}, \ket{\tilde{\mathbf{x}^*}} \bra{\tilde{\mathbf{x}^*}} \otimes \sigma_E) \geq 1 - \epsilon_T'
\end{equation}
Hence, the overall state before the key generation measurement is approximately a separate state if we can assure that the state on $\mathcal{K}^{\otimes N}$ is close to a $X$ eigenstate, i.e. we can determine its $X$-basis measurement result. Note that the state $\ket{\tilde{\mathbf{x}^*}} \bra{\tilde{\mathbf{x}^*}} \otimes \sigma_E$ yields $\rho_{KE}^{ideal}$ after $Z$-basis measurements and privacy amplification, and fidelity never decreases after quantum operations. Hence, the squashing protocol is secure (and so is the actual protocol) as long as we can determine the X-basis measurement result of the key generation system $\mathcal{K}^{\otimes N}$.

In order to gain information of the $X$-basis measurement result of $\mathcal{K}^{\otimes N}$, we make use of the ancillary system $\mathcal{H}_R$ left after the squashing operation $\Lambda$. We measure $\mathcal{H}_R$ by $\mathcal{M}_R$ and obtain a result $\gamma$, which provides information of the $X$-basis measurement result of $\mathcal{K}^{\otimes N}$. To be more specific, given each measurement result $\gamma$ on $\mathcal{H}_R$, suppose the candidates of $\mathcal{K}^{\otimes N}$ $X$-basis measurement result are summarized in the set $T_\gamma$. Suppose the cardinality of the candidate sets, except for a small probability $\epsilon_T$, can be bounded by:
\begin{equation}
|T_\gamma| \leq d^{Ns}.
\end{equation} 
In this case, suppose we make $m = N(s + \zeta)$ random $X$-parity checks, i.e. phase-error correction (see Appendix~\ref{AppendixA3}), we can derive the $X$-basis measurement result of $\mathcal{K}^{\otimes N}$ with an exponentially small failure probability $\epsilon_T' \equiv \epsilon_T + d^{-N\zeta}$ \cite{Bennett1996Mixed}. Hence, we claim the $\sqrt{\epsilon_T'}$-secret of the following single-body phase-error correction protocol:

{\bf\underline{phase-error correction protocol}}
\begin{enumerate}
\item \textbf{State distribution} Alice and Bob share a bipartite state $\rho_{AB}$ in the space $(\mathcal{H}_A \otimes \mathcal{H}_B)^{\otimes N}$ after $N$ runs of quantum communications.
\item \textbf{Squashing} They apply $\Lambda$ on $\rho_{AB}$ and convert it to a key space $\mathcal{K}^{\otimes N}$ and an ancillary space $\mathcal{H}_R$, i.e. $\Lambda(\rho_{AB}) \in \mathcal{K}^{\otimes N} \otimes \mathcal{H}_R$. The $Z$-basis measurement statistics on $\mathcal{K}^{\otimes N}$ is the same as $\kappa_{rec}$ in the actual protocol.
\item \textbf{Ancillary measurement} They measure $\mathcal{H}_R$ by $\mathcal{M}_R$ to obtain result $\gamma$. The candidate sets cardinality $|T_\gamma| \leq d^{Ns}$ except for a small probability $\epsilon_T$. 
\item \textbf{phase-error correction} For $m = N(s + \zeta)$, they randomly choose $m$ $N$-digit strings $\{W_j\}_{j=1,\cdots,m}$ and perform X-parity measurements $\{\mathcal{M}_X(W_j)\}$ on $\mathcal{K}^{\otimes N}$ to determine its $X$-basis measurement result.
\item \textbf{Key generation} They choose an arbitrary linearly independent set $\{V_k\}_{k=1,\cdots,N-m}$ satisfying $V_k \cdot W_j = 0$ for any ($j$,$k$). They perform $Z$-parity check measurements $\{\mathcal{M}_Z(V_k)\}$ to obtain the ($N-m$)-digit final key $\kappa_{fin}$.
\end{enumerate}

It only remains to show the equivalence of the phase-error correction protocol and the squashing protocol. Observe that this can be done if we can, just like that in Ref. \cite{koashi2009simple}, swap the key generation step with the phase-error correction step and omit the latter as well. However, this is the point where the two-dimensional proof cannot be extended trivially to higher dimensions. In the two-dimensional proof, $X$ and $Z$ operators are also observables, and thus the parity check measurements have decent observable forms. In this case, the commuting argument is guaranteed by the commutation of $X$-parity check observables and $Z$-parity check observables. However, in high dimensions, the parity check measurements with multiple outcomes cannot be easily expressed as Pauli operators, so here we describe the parity check measurements with measurement (Kraus) operators. In the section below, we show that this commuting argument is still valid: as long as $W_j$ is orthogonal with $V_k$, the statistics of the $Z$-parity measurement $\{\mathcal{M}_Z(V_k)\}$ would not change even if we perform in prior an $X$-parity measurement $\{\mathcal{M}_X(W_j)\}$ \eqref{swapping}. In this way, we manage to show the security of the actual protocol:
\begin{theorem}
If the actual protocol can be converted into a squashing protocol with squashing operation $\Lambda$ and ancillary measurement $\mathcal{M}_R$ such that:
\begin{enumerate}
\item The $Z$-basis measurement statistics on $\mathcal{K}^{\otimes N}$ is the same as $\kappa_{rec}$ in the actual protocol.
\item Given each measurement outcome $\gamma$ on $\mathcal{H}_R$, the size of $X$-basis measurement outcome on $\mathcal{K}^{\otimes_N}$ is bounded by $|T_\gamma| \leq d^{Ns}$, except for a small probability $\epsilon_T$,
\end{enumerate}
then the squashing protocol is $\sqrt{\epsilon_T'}$-secret, and the actual protocol is  $(\sqrt{\epsilon_T'} + \epsilon_{cor})$-secure, where $\epsilon_T' = \epsilon_T + d^{-N\zeta}$ and $m = N(s + \zeta)$.
\end{theorem} 

It is useful to derive the key-rate formula based on phase error of high-dimensional QKD. Our goal is to determine the X-basis measurement outcome $X^*$ on $\mathcal{K}^{\otimes_N}$, and we infer $X^*$ based on the ancillary measurement result $\gamma$. Based on each $\gamma$, suppose we make an estimation of $X^*$ as $X_\gamma$. Denote the phase error number vector of a given $\gamma$ as $\vec{N}_{ph} := \vec{wt}(X_\gamma-X^*)$, where the subtraction is defined for GF($d$) strings, and the vector weight function for GF($d$) is defined as:
\begin{equation}
 \vec{wt}(\mathbf{a}) = 
 \begin{bmatrix}
 \text{No. of 0 in $\mathbf{a}$}\\
 \text{No. of 1 in $\mathbf{a}$}\\
 \cdots\\
  \text{No. of $d$-1 in $\mathbf{a}$}
 \end{bmatrix}
\end{equation}
Hence, the phase error number vector $\vec{N}_{ph}$ counts the numbers of different types of phase error of our estimation $X_\gamma$. Denote the average phase error number vector for all $\gamma$ as $\vec{N}_{ph}$, and the phase-error rate vector as $\vec{E}_{ph} := \vec{N}_{ph}/N$, i.e. it counts the phase-error rate of different types of phase error. Based on Shannon's typical sequences arguments, taking the reconciliation cost as $l_{ec}$ bits, the key generation length of a $d$-dimensional QKD is
\begin{equation}
\begin{aligned}
R &= N -m-l_{ec}/\log_2 d \text{ (dits)}\\
&\leq N(1-H_d(\vec{E}_{ph})) - l_{ec}/\log_2 d \text{ (dits)}\\
&= N(\log_2d- H_2(\vec{E}_{ph})) - l_{ec} \text{ (bits)},
\end{aligned}
\end{equation}
where $H_2$ and $H_d$ are the $\log_2$ and $\log_d$ based Shannon-entropy functions respectively.

\subsection{The commuting argument in high dimension}
Denote $N$ as the number of digits of the raw key, i.e. the rounds of quantum communication. $d$ is the dimension of the primitive Hilbert space $\mathcal{H}_d$, i.e. we are operating on qudits. We confine $d$ to be a prime power, i.e. $d = p^r$, where $p$ is a prime number, so that we can define the finite field GF($d$). Denote $\gamma_p$ as the complex number satisfying $\gamma_p^p = 1$. In the arguments below, the addition, multiplication and dot product are to be understood within GF($d$).

Given two $N$-digit GF($d$) strings $\mathbf{a}$ and $\mathbf{b}$ such that $\mathbf{a}\cdot\mathbf{b} = 0$, we would like to show that 
\begin{equation}
\label{swapping}
\mathcal{M}^{\mathbf{a}}_{Z} \circ \mathcal{M}^{\mathbf{b}}_{X} = \mathcal{M}^{\mathbf{a}}_{Z} \text{ in terms of measurement statistics.}
\end{equation}
If $\mathbf{a} = 0$, the argument follows trivially as the measurement result is always zero. For non-zero $\mathbf{a}$ and an arbitrary state $\rho$, the probability that it falls into the $l$-th eigenspace of $\mathcal{M}_{Z} (\mathbf{a})$ is:
\begin{equation}
\label{ProbZ}
\sum_{\mathbf{z}\cdot\mathbf{a} = l}\bra{\mathbf{z}}\rho\ket{\mathbf{z}}
\end{equation}

On the other hand, the state after $\mathcal{M}_{X} (\mathbf{b})$ is:
\begin{equation}
\sum_{j=0}^{d-1}\sum_{\mathbf{x_j}\text{,}\mathbf{x_j'}\cdot\mathbf{b}=j}\ket{\tilde{\mathbf{x_j}}}\bra{\tilde{\mathbf{x_j}}}\rho\ket{\tilde{\mathbf{x_j'}}}\bra{\tilde{\mathbf{x_j'}}}
\end{equation}
The probability that its $\mathcal{M}_{Z} (\mathbf{a})$ result falls into the $l$-th eigenspace is thus:
\begin{equation}
\label{ProbXZ}
\sum_{\mathbf{z}\cdot\mathbf{a} = l}\bra{\mathbf{z}}\left(\sum_{j=0}^{d-1}\sum_{\mathbf{x_j}\text{,}\mathbf{x_j'}\cdot\mathbf{b}=j}\ket{\tilde{\mathbf{x_j}}}\bra{\tilde{\mathbf{x_j}}}\rho\ket{\tilde{\mathbf{x_j'}}}\bra{\tilde{\mathbf{x_j'}}}\right)\ket{\mathbf{z}}
\end{equation}
Our task is to show that \eqref{ProbZ} = \eqref{ProbXZ}. 

We first examine three lemmas. In the argument below, we ignore the scaling constants to simplify the notations. 
\begin{lemma}
\label{summation1}
\hfill
\begin{equation}
\sum_{\mathbf{z}}\gamma_p^{\mathbf{z}\cdot\mathbf{x}} = 
\begin{cases}
1 \quad &\mathbf{x} = \mathbf{0}\\
0 \quad &\mathbf{x} \neq \mathbf{0} 
\end{cases}
\end{equation}
where $\mathbf{z}$ traverses all GF(d) strings with some fixed length.  
\end{lemma}

\begin{lemma}
\label{summation2}
If $\mathbf{x}$ is 0 at one of the non-zero positions of $\mathbf{a}$, then for any GF(d) member $l$:
\begin{equation}
\sum_{\mathbf{z}\cdot\mathbf{a} = l}\gamma_p^{\mathbf{z}\cdot\mathbf{x}} = 
\begin{cases}
1 \quad &\mathbf{x} = \mathbf{0}\\
0 \quad &\mathbf{x} \neq \mathbf{0} 
\end{cases} 
\end{equation}
\end{lemma}
{\bf Proof:}
Since $\mathbf{x}$ is 0 at one of the non-zero positions of $\mathbf{a}$, that digit is essentially redundant in the summation. Denote the ($N-1$)-digit sub-string of $\mathbf{z}$ with that digit removed as $\mathbf{z'}$. Since $\mathbf{z}$ traverses all $N$-digit strings that satisfy $\mathbf{z}\cdot\mathbf{a} = l$, $\mathbf{z'}$ actually takes values of all ($N-1$)-digit strings. To see this, observe that for any ($N-1$)-digit string $\mathbf{z'}$, there is one and only one $N$-digit string $\mathbf{z}$ that satisfies $\mathbf{z}\cdot\mathbf{a} = l$ corresponds to it. This is guaranteed as we are working with a field structure. Hence, we transformed the summation to the case of Lemma \ref{summation1}.

\begin{lemma}
\label{summation3}
\hfill
\begin{equation}
\sum_{\mathbf{z}\cdot\mathbf{a} = l}\gamma_p^{\mathbf{z}\cdot\mathbf{x}} = 
\begin{cases}
\gamma_p^{x_0 l} \quad &\mathbf{x} = x_0\mathbf{a}, \quad x_0 = 0,1,\cdots,d-1\\
0 \quad &\text{otherwise} 
\end{cases} 
\end{equation}
\end{lemma}
{\bf Proof:}
Assume $\mathbf{a}$ is non-zero at digit $n$. As we are working with a field structure, there always exists $x_0 \in \{0,1,\cdots,d-1\}$ such that $\mathbf{x}_n = x_0\mathbf{a}_n$. We make the following decomposition:
\begin{equation}
\sum_{\mathbf{z}\cdot\mathbf{a} = l}\gamma_p^{\mathbf{z}\cdot\mathbf{x}} = \sum_{\mathbf{z}\cdot\mathbf{a} = l}\gamma_p^{x_0(\mathbf{z}\cdot\mathbf{a})}\gamma_p^{\mathbf{z}\cdot(\mathbf{x}-x_0\mathbf{a})} = \gamma_p^{x_0 l}\sum_{\mathbf{z}\cdot\mathbf{a} = l}\gamma_p^{\mathbf{z}\cdot(\mathbf{x}-x_0\mathbf{a})}
\end{equation}
Note that $(\mathbf{x}-x_0\mathbf{a})$ is guaranteed to be zero at digit $n$, where $\mathbf{a}$ is non-zero. We can then apply Lemma \ref{summation2} to arrive at the desired result.

We are ready to prove the main claim that \eqref{ProbZ} = \eqref{ProbXZ}:
\begin{equation}
\begin{aligned}
&\sum_{\mathbf{z}\cdot\mathbf{a} = l}\bra{\mathbf{z}}\left(\sum_{j=0}^{d-1}\sum_{\mathbf{x_j}\text{,}\mathbf{x_j'}\cdot\mathbf{b}=j}\ket{\tilde{\mathbf{x_j}}}\bra{\tilde{\mathbf{x_j}}}\rho\ket{\tilde{\mathbf{x_j'}}}\bra{\tilde{\mathbf{x_j'}}}\right)\ket{\mathbf{z}}\\
= &\sum_{\mathbf{z}\cdot\mathbf{a} = l}\sum_{j=0}^{d-1}\sum_{\mathbf{x_j}\text{,}\mathbf{x_j'}\cdot\mathbf{b}=j}\bra{\tilde{\mathbf{x_j}}}\rho\ket{\tilde{\mathbf{x_j'}}}\gamma_p^{-\mathbf{z}\cdot\mathbf{x_j}+\mathbf{z}\cdot\mathbf{x_j'}}\\
= &\sum_{\mathbf{z}\cdot\mathbf{a} = l}\sum_{j=0}^{d-1}\sum_{\mathbf{x_j}\text{,}\mathbf{x_j'}\cdot\mathbf{b}=j}\sum_{\mathbf{k}\text{,}\mathbf{k'}}\bra{\mathbf{k}}\rho\ket{\mathbf{k'}}\gamma_p^{-\mathbf{z}\cdot\mathbf{x_j}+\mathbf{z}\cdot\mathbf{x_j'}-\mathbf{x_j'}\cdot\mathbf{k'}+\mathbf{x_j}\cdot\mathbf{k}}\\
= &\sum_{\mathbf{k}\text{,}\mathbf{k'}}\sum_{j=0}^{d-1}\sum_{\mathbf{x_j}\text{,}\mathbf{x_j'}\cdot\mathbf{b}=j}\bra{\mathbf{k}}\rho\ket{\mathbf{k'}}\gamma_p^{-\mathbf{x_j'}\cdot\mathbf{k'}+\mathbf{x_j}\cdot\mathbf{k}}\sum_{\mathbf{z}\cdot\mathbf{a} = l}\gamma_p^{\mathbf{z}\cdot(\mathbf{x_j'}-\mathbf{x_j})}\\
= &\sum_{\mathbf{k}\text{,}\mathbf{k'}}\sum_{j=0}^{d-1}\sum_{x_0=0}^{d-1}\sum_{\substack{\mathbf{x_j}\text{,}\mathbf{x_j'}\cdot\mathbf{b}=j \\ \mathbf{x_j'}-\mathbf{x_j}=x_0\mathbf{a}}}\bra{\mathbf{k}}\rho\ket{\mathbf{k'}}\gamma_p^{x_0 l}\gamma_p^{\mathbf{x_j}\cdot(\mathbf{k}-\mathbf{k'})}\gamma_p^{-x_0 \mathbf{k'}\cdot\mathbf{a}}\\
= &\sum_{\mathbf{k}\text{,}\mathbf{k'}}\sum_{x_0=0}^{d-1}\sum_{\mathbf{x_j}}\bra{\mathbf{k}}\rho\ket{\mathbf{k'}}\gamma_p^{x_0( l- \mathbf{k'}\cdot\mathbf{a})}\gamma_p^{\mathbf{x_j}\cdot(\mathbf{k}-\mathbf{k'})}\\
= &\sum_{\mathbf{k'}\cdot\mathbf{a} = l}\bra{\mathbf{k}}\rho\ket{\mathbf{k'}}\delta_{\mathbf{k} = \mathbf{k'}}\\
= &\sum_{\mathbf{k}\cdot\mathbf{a} = l}\bra{\mathbf{k}}\rho\ket{\mathbf{k}},
\end{aligned}
\end{equation}
which is exactly the $\mathcal{M}^{\mathbf{a}}_{Z}$ statistics without $\mathcal{M}^{\mathbf{b}}_{X}$ in \eqref{ProbZ}. Lemma~\ref{summation1} and Lemma~\ref{summation3} are applied in the 6-th and 4-th equalities. Notice that in the summation we require $\mathbf{x_j}\text{,}\mathbf{x_j'}\cdot\mathbf{b}=j$ for any GF($d$) member $j$, which is equivalent to $(\mathbf{x_j'}-\mathbf{x_j})\cdot\mathbf{b}=0$. Hence, $\mathbf{x_j'}-\mathbf{x_j} = x_0\mathbf{a}$ is within the summation range since we picked $\mathbf{a}\cdot\mathbf{b} = 0$. Therefore, the swapping argument that performing $X$ parity checks and then $Z$ hashed key generation is equivalent to the latter on its own can be extended to higher-dimensional cases.

\section{Symmetry-based security analysis of high-dimensional PM QKD}
\label{Appendix3}

\subsection{symmetric encoding protocols}
Based on the security proof of high-dimensional QKD developed above and the symmetric encoding security analysis of PM QKD \cite{Zeng2019Symmetryprotected}, we provide the security analysis of $d$-dimensional PM QKD where $d$ is a prime number. We introduce the entanglement-based symmetric encoding QKD protocol, as shown in Fig.~ \ref{EBPMQKD} below. Alice and Bob share the state $\rho_{AB}$ on system $A$ and $B$, and each holds an ancillary $d$-dimensional qudit system $A'$ and $B'$ initially on the state $\ket{+}_d := \frac{1}{\sqrt{d}}\sum_{j=0}^{d-1} \ket{j}$.    

\begin{figure}[htbp!] 
\centering    
\captionsetup{justification = raggedright}
\includegraphics[width=0.45\textwidth]{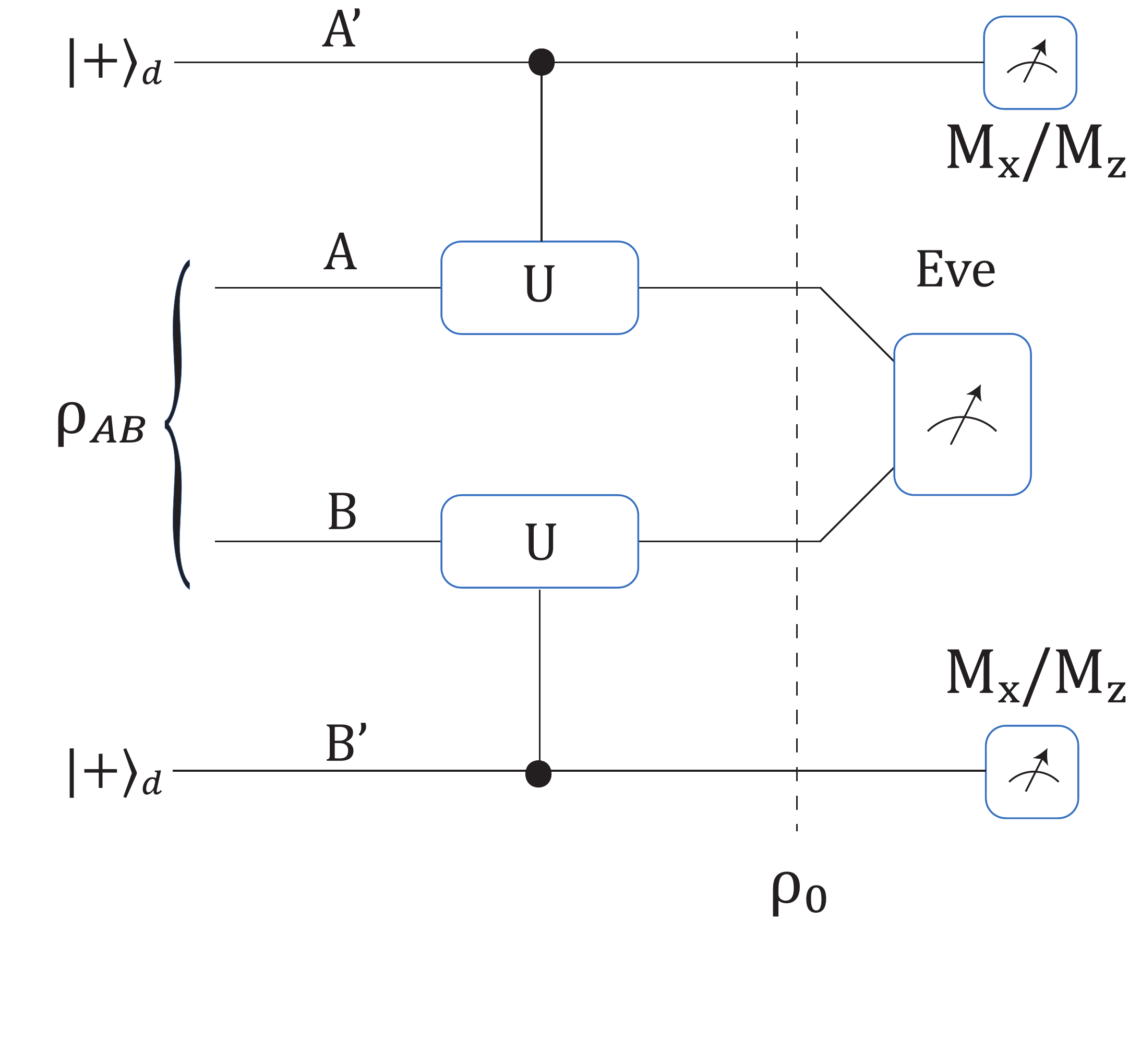}
\caption{Schematic diagram of the entanglement-based $d$-dimensional PM QKD, where $\rho_{AB}$ is a bipartite state on two optical modes, and the encoding operation $U$ rotates the coherent state by $2\pi/d$. The optical mode is phase-rotated by $2\pi k/d$ if the $k$-th control dit is triggered. Same as Fig.~\ref{EBSymmetric} in the main text.}
\label{EBPMQKD}
\end{figure}

Alice and Bob send the shared state $\rho_{AB}$ through a controlled encoding operation, where 
\begin{equation}
C_{A'A}(U) = \sum_{j = 0}^{d-1} \ket{j}_{A'}\bra{j} \otimes U_A^j,
\end{equation}
and similarly for $C_{B'B}(U)$. The unitary encoding operation is $d$-rotation symmetric, i.e., $U^d = I$. They then send the systems $A$ and $B$ further to Eve for detection. At the end of the quantum communications, they share $N$ pairs of qudit systems for key generation. 

Following the security proof of high-dimensional QKD given above, taking $A'$ as the key generation system and $B'$ as the ancillary system, we need to determine the X-measurement results of $A'$ with the knowledge of that of $B'$. This can be done as long as the originally separate $\ket{+}_{A'}$ and $\ket{+}_{B'}$ are now entangled after the symmetric encoding operations. In other words, we need the shared state $\rho_{AB}$ to give the same encoded state after different encoding operations, i.e. $\rho_{AB}$ being the eigenstate of $U_A \otimes U_B$.  

Since we have $(U\otimes U)^d = I$, the eigenvalues of $(U \otimes U)$ are $\{ \gamma_d^j := \exp(i\frac{2\pi}{d} l) \}_{l=0}^{d-1}$. The eigenspace of $\gamma_d^l$ is denoted by $\mathcal{H}^{(l)}$. Denote a generic state $\ket{\psi}\in \mathcal{H}^{(l)}$ as $\ket{\psi_l}$, hence
\begin{equation}
(U\otimes U)\ket{\psi_l} = \gamma_d^l \ket{\psi_l}.
\end{equation}

{\bf \underline{High-dimensional symmetric encoding protocol}}
\begin{enumerate}
\item \textbf{State preparation}: Alice and Bob share a state $\rho_{AB}$ at the beginning of each run. They initialize their qudits $A'$ and $B'$ in $\ket{+}_d$. They apply the control gate $C_{A'A}(U)$ and $C_{B'B}(U)$ respectively.
\item \textbf{Measurement}: Alice and Bob send $\sigma_{AB}$ to an untrusted party, Eve, who is supposed to perform joint measurement and announce the detection results. 
\item \textbf{Sifting}: Given a specific announcement of Eve, Alice and Bob keep or discard the qudits of systems $A'$ and $B'$. Alice and Bob perform the above steps for many rounds and end up with a joint $2N$-qudits state $\rho_{A'B'} \in (\mathcal{H}_A' \otimes \mathcal{H}_B')^{\otimes N}$.
\item \textbf{Key generation}: Alice and Bob perform local $Z$-measurements on $\rho_{A'B'}$ to obtain two correlated raw key strings $\kappa_A$ and $\kappa_B$. They reconcile the key string to $\kappa_{rec}$ by an encrypted classical channel, consuming $l_{ec}$-bit keys.
\end{enumerate}

We first consider the case when a $l$-symmetric state $\ket{\psi_l}_{AB}$ is the input state of the entanglement-based protocol. The initial state is
\begin{equation}
\begin{aligned}
\ket{++}_{A',B'} \ket{\psi_l}_{A,B} &= \frac{1}{d}\sum_{j,k=0}^{d-1} \ket{jk}_{A',B'} \ket{\psi_l}_{A,B} \\
&= \frac{1}{\sqrt{d}} \sum_{u=0}^{d-1} \ket{\Phi_{u,0}}_{A',B'} \ket{\psi_l}_{A,B}. \\
\end{aligned}
\end{equation}

After the encoding unitary operation, $C_{A'A}(U)$ and $C_{B'B}(U)$, the state becomes
\begin{equation}
\ket{\Psi}_{A',B',A,B} = \frac{1}{\sqrt{d}} \sum_{u=0}^{d-1} \ket{\Phi_{u,l}}_{A',B'} \ket{\psi_l^u}_{A,B},
\label{Psi}
\end{equation}
here $\ket{\psi_l^u}:= (I\otimes U^u) \ket{\psi_l}$. To derive Eq.~ \eqref{Psi}, we have applied the following property
\begin{widetext}
\begin{equation}
[C_{A'A}(U) \otimes C_{B'B}(U)] \ket{\Phi_{u,v}}_{A',B'} \ket{\psi_l}_{A,B}  =  \ket{\Phi_{u,v+l}}_{A',B'} \ket{\psi_l^u}_{A,B}.
\end{equation}
\end{widetext}

In this case, the space of $A1, B1$ is spanned by $\{\Phi_{u,l}\}_{u=0}^{d-1}$. Note that
\begin{equation}
\tr\left[\Phi_{u,l} \ket{\tilde{j},\tilde{k}}_{A1,B1}\bra{\tilde{j},\tilde{k}}\right]= \frac{1}{d} \delta_{j,l-k},
\end{equation}
which is irrelevant of $u$. Therefore, if Alice and Bob perform $X$-measurement on $A'$ obtaining result $l_a$, the $X$-measurement result $l_b$ is directly related as $l_a = l - l_b \text{ mod }d$. This implies that the protocol is completely secure as long as Alice and Bob share $l$-symmetric states for a fixed $l$. Hence, we arrive at the security of the prime-dimensional symmetric encoding QKD for symmetric states:
\begin{theorem}
In the prime-dimensional symmetric encoding protocol, the X-measurement result on $A'$ can be determined exactly with the X-measurement result on $B'$ if Alice and Bob share a mixture of $l$-symmetric states, for fixed $l$, at the beginning of each run. Hence, in that case, the protocol is completely secure.
\end{theorem}

However, in the general setup, the shared state $\rho_{AB}$ is usually not a mixture of pure symmetric states, but a mixture of different symmetric states, that is,
\begin{equation}
\label{mixture of parity}
\rho_{AB} = \sum_{l = 0}^{d-1}\sum_{j} p_l^{(j)}\ket{\psi_l^{(j)}}\bra{\psi_l^{(j)}},
\end{equation}
where $\ket{\psi_l^{(j)}}$ are the $l$-symmetric states and $\sum_{l = 0}^{d-1}\sum_{j} p_l^{(j)} = 1$. This mixture source is equivalent to Alice and Bob preparing $l$-symmetric states for probability of $\sum_{j} p_l^{(j)}$ for each run. However, the parity information, i.e. which symmetric state is sent each round, is not known to Alice and Bob (and known by Eve in the worst case scenario). Hence, they cannot deal with each symmetric state separately, and thus there is no longer perfect privacy. We define the phase-error rate vector as
\begin{equation}
\vec{E}_{ph} = \left[\frac{N_0}{N},\frac{N_1}{N},\cdots,\frac{N_{d-1}}{N}\right],
\end{equation}  
where $N_l$ is the number of detections caused by $l$-symmetric states. According to the key-rate formula of high-dimensional QKD, the asymptotic key rate of the $d$-dimensional symmetric encoding protocol is
\begin{equation}
\label{key-rate formula}
r = \log_2d- H_2(\vec{E}_{bit}) - H_2(\vec{E}_{ph}) \text{ bits}
\end{equation}

\subsection{High-dimensional PM QKD with continuous randomization}
\label{AppendixC2}
The high-dimensional entanglement-based PM QKD given below falls into the category of symmetric encoding protocol discussed above, and hence its key rate is given by Eq.~\eqref{key-rate formula}. The encoding operation $U$ is given by
\begin{equation}
U = e^{i \frac{2\pi}{d} a^\dagger a},
\end{equation}
where $a$ is the annihilation operator. It is clear that $U$ is $d$-rotation symmetric, i.e. $U^d = I$. It can be seen that, when applied on the Fock state $\ket{n}$, this operation adds an additional phase $e^{\frac{2\pi i}{d}n}$. Hence, we intend to generate mixture of Fock states as input through continuous randomization.

\underline{\textbf{High-dimensional entanglement-based PM QKD}}

\underline{\textbf{with continuous randomization}}
\begin{enumerate}
\item \textbf{State preparation}: Alice and Bob prepare the coherent state $\ket{\sqrt{\mu/2}~e^{i\phi_a}}_A \otimes \ket{\sqrt{\mu/2}~e^{i\phi_b}}_B$ on two optical modes $A$ and $B$, where $\phi_a$ and $\phi_b$ are selected randomly from $[0,2\pi)$, and $\mu$ taken from multiple values as in decoy methods. They initialize their qudits $A'$ and $B'$ in $\ket{+}_d := \frac{1}{\sqrt{d}}\sum_{j=0}^{d-1} \ket{j}$. They apply the control gate $C_{A'A}(U)$ and $C_{B'B}(U)$ respectively, where $U$ rotates a coherent state by $2\pi/d$.
\item \textbf{Measurement}: Alice and Bob send the two optical modes $AB$ to an untrusted party, Eve, who is supposed to perform joint measurement and announce the detection results: no-click, double-click, $L$-click or $R$-click. 
\item \textbf{Sifting}: After many rounds of quantum communications, Alice and Bob keep only the rounds with $L$ or $R$ click. They announce the random phases $\phi_a$ and $\phi_b$ publicly and keep only the rounds where $|\phi_a - \phi_b| \in \{k\frac{2\pi}{d}\}_{k=0}^{d-1}$. They end up with a joint $2N$-qudits state $\rho_{A'B'} \in (\mathcal{H}_A' \otimes \mathcal{H}_B')^{\otimes N}$.
\item \textbf{Parameter estimation}: Alice and Bob estimate the phase-error vector $\vec{E}_{ph}$ with decoy states.
\item \textbf{Key generation}: Alice and Bob perform local $Z$-measurements on $\rho_{A'B'}$ to obtain two correlated raw key strings $\kappa_A$ and $\kappa_B$. They reconcile the key string to $\kappa_{rec}$ by an encrypted classical channel, consuming $l_{ec}$-bit keys. They perform privacy amplification according to the phase-error vector to generate the final keys.
\end{enumerate}

For states with $\phi_a - \phi_b = \delta$, the continuous randomization in fact generates the input state:
\begin{widetext}
\begin{equation}
\frac{1}{2\pi} \int_0^{2\pi} d\phi \ket{\sqrt{\mu/2}~e^{i\phi}}_A\bra{\sqrt{\mu/2}~e^{i\phi}} \otimes \ket{\sqrt{\mu/2}~e^{i(\phi+\delta)}}_B\bra{\sqrt{\mu/2}~e^{i(\phi+\delta)}} = \sum_{k=0}^{\infty} P_{\mu}(k)\ket{\Bar{k}^\delta}_{AB}\bra{\Bar{k}^\delta},
\end{equation}
\end{widetext}
where $P_{\mu}(k) = e^{-\mu}\frac{\mu^k}{k!}$ is the Poisson distribution. The $k$-photon state $\ket{\Bar{k}^\delta}_{AB}$ is
\begin{equation}
\label{kphoton}
\ket{\Bar{k}^\delta}_{AB} = \frac{(a^\dagger + e^{i\delta}b^\dagger)^k}{\sqrt{2^k k!}}\ket{00}_{AB},
\end{equation}
which is a $k$-symmetric state. We can thus define the phase-error rate vector with entries:
\begin{equation}
\vec{E}_{ph}(k) = \sum_{n=0}^{\infty} q_{nd+k},\quad k\in\{0,\cdots,d-1\},
\end{equation}
where $q_k$ is the fraction of detection events caused by $\ket{\Bar{k}^\delta}_{AB}$.

Since Fock states are invariant with intensity $\mu$, their yields do not vary with $\mu$, and we can apply the decoy methods given the overall gain $Q_\mu$ \cite{Lo2005Decoy,Ma2005practical}:
\begin{equation}
Q_\mu = \sum_{k=0}^{\infty} P_\mu(k)Y_k,
\end{equation}
and the fraction of detection is given by
\begin{equation}
\label{Eq:Eph calculation}
q^\mu_k = P_\mu(k) \frac{Y_k}{Q_\mu}.
\end{equation}

\subsection{High-dimensional PM QKD with discrete randomization}
\label{Appendix:discrete randomization}
It is a common practice to approximate the ideal continuous randomization with discrete randomization \cite{Lo2005Decoy,Cao2015discrete,Ma2018phase}. In the state preparation stage of the $d$-dimensional entanglement-based PM QKD, instead of continuously randomizing the phase of the coherent states, Alice and Bob apply a $D$-slice discrete phase randomization for a large $D$, and postselect phase-matched rounds similarly.

For the rounds where Alice and Bob share a phase reference difference of $\delta$, they generate the input state as a mixture of ``pseudo"-Fock states:
\begin{widetext}
\begin{equation}
\label{pseudo Fock states}
\frac{1}{D}\sum_{j=0}^{D-1} \ket{\sqrt{\mu/2}~e^{i\frac{2\pi j}{D}}}_A \bra{\sqrt{\mu/2}~e^{i\frac{2\pi j}{D}}} \otimes \ket{\sqrt{\mu/2}~e^{i(\frac{2\pi j}{D}+\delta)}}_B \bra{\sqrt{\mu/2}~e^{i(\frac{2\pi j}{D}+\delta)}} = \sum_{k=0}^{\infty} P^{\mu}_D(k)\ket{\Bar{\lambda}_k^\delta}_{AB}\bra{\Bar{\lambda}_k^\delta}
\end{equation}  
\end{widetext}
where
\begin{equation}
\begin{aligned}
&\ket{\Bar{\lambda}_k^\delta} = \frac{e^{-\mu/2}}{\sqrt{P_{\mu}(k)}}\sum_{n=0}^{\infty} \frac{(\sqrt{\mu})^{nD+k}}{\sqrt{(nD+k)!}}\ket{\overline{nD+k}^\delta}\\
&P_D^{\mu}(k) = \sum_{n=0}^{\infty}\frac{\mu^{nd+k} e^{-\mu}}{(nd+k)!},
\end{aligned}
\end{equation}
with $k$-photon state $\ket{\Bar{k}^\delta}$ defined in Eq.~\eqref{kphoton}.

The $k$-pseudo Fock state $\ket{\Bar{\lambda}_k^\delta}$ is also a $k$-symmetric state of $U \otimes U$, so the security analysis still applies. It is however less favoured than Fock states since for moderate $D$ it varies with intensity $\mu$, thus enabling Eve to discriminate signal states with decoy states, cracking the decoy method \cite{Lo2005Decoy}. We therefore require $D$ to be large so that the yield of $\ket{\Bar{\lambda}_k^\delta}$ approximates the yield of $\ket{\Bar{k}^\delta}$, which is invariant with intensity. Denote the yield and the detection fraction of the non-ideal $k$-th symmetric state as $Y_{\lambda_k}$ and $q_{\lambda_k}$. In Ref. \cite{Zeng2019Symmetryprotected}, a bound between the deviation of $Y_{\lambda_1}$ and $q_{\lambda_1}$ from $Y_1$ and $q_1$ is given, and can be straightforwardly extended to general $k$-photon states:

\begin{equation}
\begin{aligned}
\label{approximate continuous}
&|Y_k - Y^{\mu}_{\lambda_k}| \leq \sqrt{\frac{\mu^D k!}{(D+k)!}}\\
&|q^{\mu}_k - q^{\mu}_{\lambda_k}| \leq \frac{\mu^{D/2+k}e^{-\mu}}{Q_\mu\sqrt{(D+k)!/k!}},
\end{aligned}
\end{equation}

A straightforward calculation reveals that Eq.~\eqref{approximate continuous} gives a tighter bound for multi-photon fractions than single-photon fraction. Hence it is sufficient to check the accuracy of single-photon fraction estimation. Denote the transmittance from Alice or Bob to Eve as $\eta$. In the first-order limit where the gain $Q_\mu \approx \eta \mu$ and yield $Y_l \approx l\eta$, Table~\ref{Estimation} below illustrates the estimation inaccuracy of single photon components in terms of $|q^{\mu}_1 - q^{\mu}_{\lambda_1}|/q^{\mu}_1$ at transmittance $\eta = 10^{-6}$ for 8 to 16 phase slices. The light intensity $\mu$ is taken as 0.1, which is a moderate value around the optimal values given in the simulations in Section~\ref{4Comparison}. The $10^{-6}$ transmittance is chosen since PM QKD can reach at most around 500 km for a $-0.2$ dB/km attenuating fiber and $20\%$ detectors. The minimum transmittance from Alice to Eve is therefore:
\begin{equation}
\eta = 10^{-0.2*250/10} \times 0.2 = 2\times10^{-6}
\end{equation}

From Table~\ref{Estimation}, it can be seen that more than 10-phase randomization is sufficient for an estimation of single-photon fraction with less than $10^{-3}$ inaccuracy. The 16-phase randomization in the original two-dimensional PM QKD is conservative.    

\begin{table}[htbp!]
\caption{Estimation inaccuracy of single-photon fraction with discrete randomization at $\eta = 10^{-6}$}
\centering
\begin{tabular}{| c |c | c| c| c| c|}
\hline
\hline
 & $D = 8$ & $D = 10$ & $D = 12$ & $D = 14$ & $D = 16$\\
\hline
$\Delta q_1/q_1$ & $0.17$ & $1.6\times10^{-3}$ & $1.3\times10^{-5}$& $8.7\times10^{-8}$ & $5.3\times10^{-10}$\\
\hline
\hline
\end{tabular}
\label{Estimation}
\end{table}

The final key-rate formula can therefore be expressed as:

\begin{equation}
r = \frac{d}{D}Q_\mu[\log_2d-H_2(\vec{E}^\mu_{bit})-H_2(\vec{q}_{\mu})],
\end{equation}
where all the parameters can be retrieved from experiments.

\section{Simulation formulae of high-dimensional PM QKD}
\label{App:SimFormula}

We present the formulae used to simulate the key rate performance of high-dimensional PM QKD in Fig.~\ref{Sim} and~\ref{Fig:Fluc}. The channel is assumed to be pure-loss and symmetric for Alice and Bob with transmittance $\eta$ (with detector efficiency taken into account). The single-photon detectors have dark count rate $p_d$. The calculations below are for single $L$-click events, and can be easily altered for $R$-click events.  

To calculate the bit-error rate vector $\vec{E}^\mu_{bit}$, assume Alice and Bob send coherent states of amplitude $\mu/2$ with phase difference $\phi + \delta$, where $\phi$ is the encoding difference and $\delta$ is the reference-frame misalignment. As computed in Ref.~\cite{Ma2018phase}, the single-click probabilities of the $L$ and $R$ detector given phase difference $\phi + \delta$ are
\begin{equation}
\begin{split}
&P_\mu^{\phi + \delta}(L) = 1-(1-p_d)\exp(-\eta \mu \cos^2((\phi + \delta)/2))\\
&P_\mu^{\phi + \delta}(R) = 1-(1-p_d)\exp(-\eta \mu \sin^2((\phi + \delta)/2)).
\end{split}
\end{equation}

Given reference misalignment $\delta$, when Alice and Bob have encoding difference $\phi_k = \frac{2\pi}{d}k$, the probability of a single $L$-click is 
\begin{equation}
P_\mu(L|\phi_k,\delta) = P_\mu^{\phi_k + \delta}(L) [1-P_\mu^{\phi_k + \delta}(R)].
\end{equation}

Since the misalignment is independent of the encoding, by the Bayesian formula, the probability of encoding difference $\phi_k$ given a single $L$-click event with misalignment $\delta$ is 
\begin{equation}
P_\mu(\phi_k|L,\delta) = \frac{P_\mu(L|\phi_k,\delta)P(\phi_k)}{Q_\mu^\delta},
\end{equation}
where $P(\phi_k) = \frac{1}{d}$ for uniform encoding. The gain $Q_\mu^\delta$ given misalignment $\delta$ can be calculated by
\begin{equation}
Q_\mu^\delta = \sum_{k=0}^{d-1}P_\mu(L|\phi_k,\delta)P(\phi_k).
\end{equation}

The $k$-th entry of the bit-error rate vector is therefore given by
\begin{equation}
\vec{E}^\mu_{bit}(k) = P_\mu(\phi_k|L) = \mathbb{E}_\delta[P_\mu(\phi_k|L,\delta)],
\end{equation}
where the expectation is taken over the distribution of misalignment $\delta$, which is deterministic for fixed misalignment and uniform for fluctuating misalignment. The total gain is the expectation
\begin{equation}
Q_\mu = \mathbb{E}_\delta[Q_\mu^\delta].
\end{equation}

To calculate the phase-error rate vector $\vec{q}_\mu$, given encoding difference $\phi_k$ and misalignment $\delta$, when Alice and Bob send the $n$-photon state, the probability of a single $L$-click is~\cite{Ma2018phase}
\begin{widetext}
\begin{equation}
P_n(L|\phi_k,\delta) = (1-p_d)(1-\eta \cos^2((\phi_k+\delta)/2))^n - (1-p_d)^2(1-\eta)^n.
\end{equation}
\end{widetext}

Averaging over the encoding, the yield of $n$-photon states under misalignment $\delta$ is given by
\begin{equation}
Y_n^\delta = \sum_{k=0}^{d-1}P_n(L|\phi_k,\delta)P(\phi_k).
\end{equation}
The total yield is therefore the expectation 
\begin{equation}
Y_n = \mathbb{E}_\delta[Y_n^\delta].
\end{equation}
We can therefore calculate the detection fraction $q_n^\mu$ of the $n$-photon states by Eq.~\eqref{Eq:Eph calculation} with yield $Y_n$ and gain $Q_\mu$. 

\end{appendix}


%

\end{document}